# Automatic assessment of text-based responses in post-secondary education: A systematic review

Rujun Gao, Hillary E. Merzdorf, Saira Anwar, M. Cynthia Hipwell, Arun Srinivasa
grj1214@tamu.edu, hmerzdorf@tamu.edu, sairaanwar@tamu.edu, cynthia.hipwell@tamu.edu, asrinivasa@tamu.edu
Texas A&M University

**Abstract**

Text-based open-ended questions in academic formative and summative assessments help students become deep learners and prepare them to understand concepts for a subsequent conceptual assessment. However, grading text-based questions, especially in large (>50 enrolled students) courses, is tedious and time-consuming for instructors. Text processing models continue progressing with the rapid development of Artificial Intelligence (AI) tools and Natural Language Processing (NLP) algorithms. Especially after breakthroughs in Large Language Models (LLM), there is immense potential to automate rapid assessment and feedback of text-based responses in education. This systematic review adopts a scientific and reproducible literature search strategy based on the PRISMA process using explicit inclusion and exclusion criteria to study text-based automatic assessment systems in post-secondary education, screening 838 papers and synthesizing 93 studies. To understand how text-based automatic assessment systems have been developed and applied in education in recent years, three research questions are considered: 1) What types of automated assessment systems can be identified using input, output, and processing framework? 2) What are the educational focus and research motivations of studies with automated assessment systems? 3) What are the reported research outcomes in automated assessment systems and the next steps for educational applications? All included studies are summarized and categorized according to a proposed comprehensive framework, including the input and output of the system, research motivation, and research outcomes, aiming to answer the research questions accordingly. Additionally, the typical studies of automated assessment systems, research methods, and application domains in these studies are investigated and summarized. This systematic review provides an overview of recent educational applications of text-based assessment systems for understanding the latest AI/NLP developments assisting in text-based assessments in higher education. Findings will particularly benefit researchers and educators incorporating LLMs such as ChatGPT into their educational activities.

## 1. Introduction

Artificial Intelligence in Education (AIEd) is an emerging field in educational technology with the potential to assist in large-scale teaching environments and give real-time feedback to students for personalized education. Although Artificial Intelligence (AI) has been incorporated into applications for more than 30 years, there is a continuing need for research to assist large-scale teaching and intelligent assistance (Zawacki-Richter, 2019). Natural Language Processing (NLP) as a subbranch of AI continues to make rapid progress in text processing, especially with the emergence of large-scale preprocessing techniques such as transformer-based models of self-attention for NLP. With the recent emergence of large language models (LLM) with powerful capabilities such as ChatGPT, there is a promising future for these applications in education (Kasneci et al., 2023). The technical foundation of automatic assessment systems has strengthened, along with an increasing number of related studies with excellent prospects for AIEd. With significant developments in AIEd, these applications can be classified into four aspects (Zawacki-Richter, 2019): 1) Decision making tools, which help in profiling and prediction for admissions decisions and course scheduling, drop-out and retention, student models, and academic achievement (Alvero et al., 2020; Chen et al., 2020; Langley, 2019); 2) Intelligent tutoring systems, which are designed for teaching course content, interacting with students, curating learning materials, facilitating collaboration, and supporting the teacher's perspective (Feng & Law, 2021; Hwang et al., 2020); 3) Adaptive systems which provide scaffolding and content personalization, support teachers to understand students' learning, use academic data to monitor and guide students, and represent knowledge in concept maps (Chen & Bai, 2010; Kabudi et al., 2021); and 4) Assessment and evaluation tools, designed for automated grading, feedback, evaluation of student understanding, engagement and academic integrity, and teaching evaluation (Huang et al., 2023; Luckin, 2017). These topic areas represent significant developments in AIEd.

Among these four aspects, assessment and evaluation tools directly relate to students' learning and outcomes. These tools are vital for students' conceptual understanding, and critical to providing timely feedback and improving student learning outcomes (Gikandi et al., 2011). In higher education, meaningful formative and summative assessment are essential to active learning. Carefully-designed assessment engages students in thinking, problem-solving, and metacognition. Such assessment helps students connect learning activities with conceptual understanding (Roselli & Brophy, 2006). Metacognition allows students to appraise their knowledge of skills and applications and to regulate their cognition through planning, monitoring, error detection, and self-evaluation (Tarricone, 2011). Metacognition and conceptual improvement are equally crucial for teachers' intentionality when reflecting on instructional practices and motivations (Gunstone & Northfield, 1994). Assessment provides teachers with real-time information about what students have learned and what remains unclear (Shepard, 2005). However, the core task of evaluating student work in quizzes and homework is either delegated to teaching assistants or converted into multiple-choice questions without feedback. Evaluating learning outcomes in a more open-ended format and providing timely feedback is challenging. High-quality assessment gauges both the teachers' intent in evaluating learning outcomes and the students' intent in providing answers to questions (Diefes‐Dux et al., 2012). In complex learning contexts, such as engineering design, instructors may assess procedural competence or high-level conceptual understanding (Cardella & Tolbert, 2014). For instruction and assessment to be both scalable and personalized, automated grading must effectively incorporate both human and machine input to meet the needs of a learning context (Geigle et al., 2016). Assessment and evaluation can be improved through AI applications to potentially empower learners with agency, collaboration, and personalization (Ouyang & Jiao, 2021).

Over the last two decades, automated scoring has been widely developed and used in content domains such as mathematics, science, and language testing (Liu et al., 2014), resulting in many papers. However, these papers have focused on automated scoring of multiple-choice questions rather than open-ended responses. Compared to multiple-choice questions, text-based open-ended questions are tedious and time-consuming for instructors to grade and do not lend themselves readily to automation, especially in ill-defined tasks requiring higher-level thinking (Wang & Brown, 2008). As a result, most automated learning management systems have focused on multiple-choice questions, with even narrower applications for science, technology, engineering, and mathematics (STEM) disciplines where most automated systems can only verify numerical answers. This is highly detrimental to instructors' mission of evaluating students on the synthesis and application of concepts and representational competence rather than on whether they obtained the correct answer (Kohl & Finkelstein, 2005).

To address the need to assess students' higher-level reasoning and competence, automated assessment systems have great potential to grow in education to analyze open-ended questions.

Given these challenges and the availability of existing literature, this study reviews recent progress in the automatic assessment of student textual responses. Our research is motivated to examine the current state of the art in text-based automatic assessment systems including automated short-answer grading, essay scoring, and writing evaluation. Moreover, this study synthesizes the current state of the literature to describe the prospect of future automatic assessment systems in post-secondary education. There are currently few relevant systematic reviews on automatic assessment systems for student texts in education. Caiza and Del Alamo (2013) conducted a relevant review of grading systems. However, their review only focused on programming assignments. A review by Efendi et al. (2022) identified emerging topic areas in automated essay scoring. However, their review lacked the broader educational impact such tools may have. Ramesh and Sanampudi (2022) only reviewed and compared current AI and machine learning (ML) techniques for automatic essay scoring with limited educational context. A review of automated writing evaluation systems by Nunes et al. (2022) included empirical educational studies with only the learning impact of writing assignments. In particular, some reviews focused on tutoring systems based on free-text context to provide guidance (Bai & Stede, 2022; Kochmar et al., 2022), which is a different research topic when compared to assessing students' short answers to specific questions. Several systematic reviews have also examined using LLMs in medical education (Kung et al., 2023; Sallam, 2023), but few have mentioned STEM fields. By addressing limitations in existing review studies, this systematic review aims to investigate up-to-date educational applications for assessing students' text-based responses. The systematic approach will ensure this paper captures the recent and rapid advances in AI and NLP, and will shed light on future post-secondary education, especially STEM education.

### 1.1. Theoretical framework for the review

We conceptualize automated assessment systems from an input-process-output (IPO) perspective (Ilgen et al., 2005) and propose a classification framework based on a comprehensive systematic review (see Figure 1). The principle is to follow a pragmatic worldview and focus on the practical consequences of the research. Input represents the textual information input into the assessment model, such as student answers, responses, or essays. Process is the text-based automated assessment system (TBAAS) for processing and analysis. Output represents the assessment results as interpreted and framed by the system, such as scores or feedback. In addition to these central elements, we identify the usage domain of each system and the influence of wider learning needs for students and instructors, the research motivations for developing the system, and the underlying educational concepts that inform its design. We also examine research outcomes beyond output as the reported system impacts on original research goals, future applications, and improvement in education. This improved IPO framework is proposed to address the different aspects of three research questions.

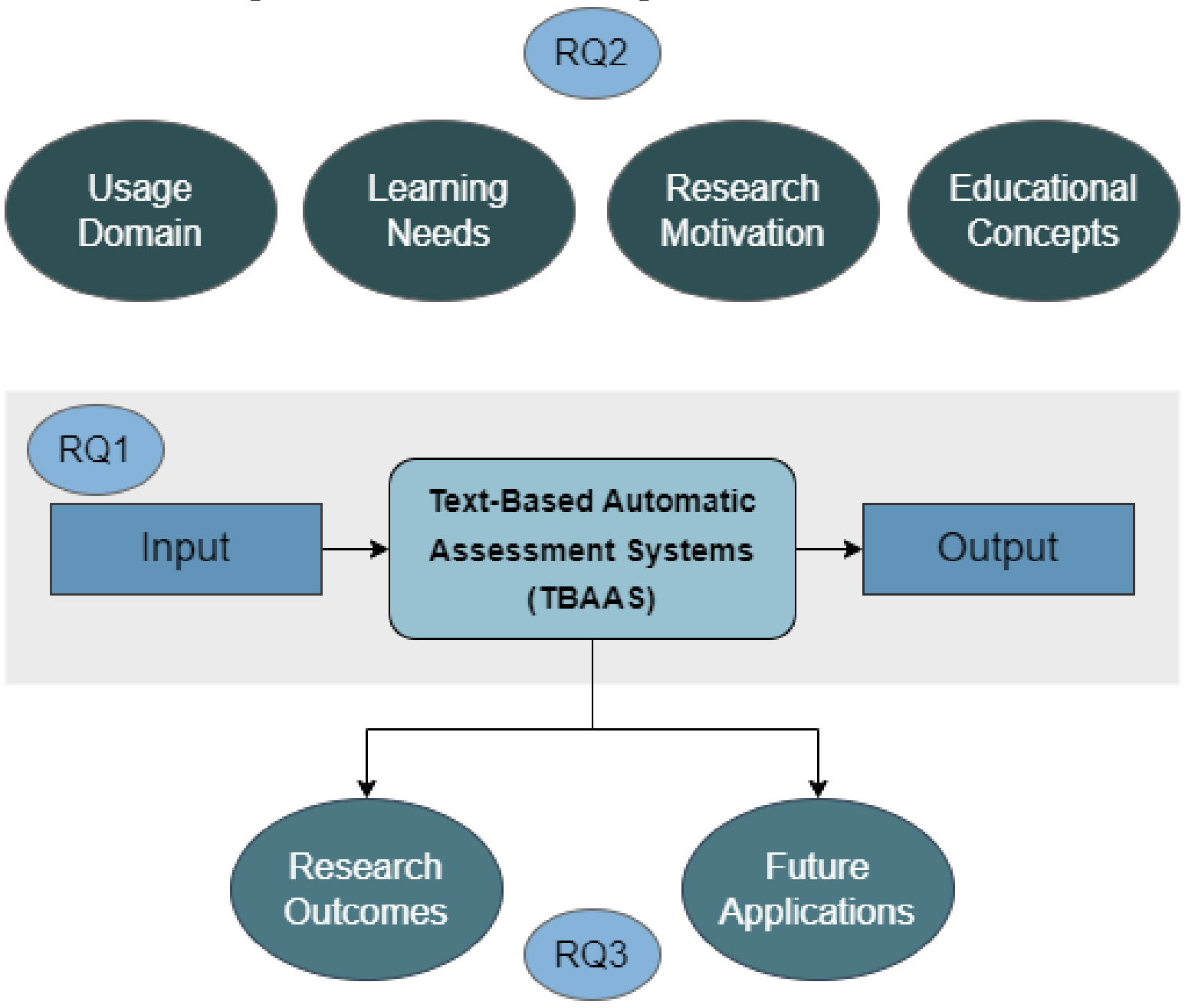

*Figure 1.* Theoretical framework for this systematic review (see Section 1.2 for research questions)

For this study, IPO guides interpretation of each study's major findings. The IPO framework guides the interpretation of our systematically reviewed results, allowing us to identify system features and how they are influenced by the usage domains, learning needs, research motivations, and educational concepts in each study.

### 1.2. Research questions

This systematic review aims to provide an overview of the current research literature on automated assessment systems in higher education. It will identify major system types and their design features in terms of input, processing, and output in post-secondary education. This review will also describe the extent to which the literature emphasizes educational benefits and integrates them with research motivations. It will lastly observe whether educational impacts are fully evaluated through research outcomes, implications, and continued applications in studies of automated assessment systems. The following research questions guide the systematic review:

- RQ1: What types of automated assessment systems can be identified using input, output, and processing framework?
- RQ2: What are the educational focuses and research motivations of studies with automated assessment systems?
- RQ3: What are the reported research outcomes in automated assessment systems, and what are the next steps for educational application?

With these research questions, we collected and synthesized information on the design features of automated assessment systems to support students and educators. From our findings, we also aim to provide insight into the current support of automated assessment systems for higher-level conceptual reasoning and deeper learning through text-based responses.

## 2. Methods

This systematic review uses a scientific and reproducible literature search strategy following the updated 2020 PRISMA (Preferred Reporting Items for Systematic Reviews and Meta-Analyses) statement (Page et al., 2021). PRISMA is a standard originating in medical research but adopted across disciplines for conducting systematic literature reviews. It is a checklist of 27 items for ensuring reproducibility and transparency, preventing bias, and following rigorous methods for collecting, screening, and synthesizing research studies (Page et al., 2021). First, based on the specific topic and the according realm, our initial search string was finalized after repeated search experiments and confirmation, and four databases were selected for an appropriate literature search. Second, a clear set of inclusion and exclusion criteria was proposed to select target journal and conference papers to answer all three research questions and exclude unrelated studies. The inclusion and exclusion criteria were applied twice. In the first screening, each paper's title and abstract were screened for all retrieved articles. For papers included in the first screening, a second round was conducted, where each paper's full text was screened. Finally, 93 papers that met the inclusion and exclusion criteria were selected, and these 93 studies were included on automatic assessment of text-based responses for this systematic review. A PRISMA flowchart (see Figure 2) was generated to summarize the whole data extraction process. Findings were then coded and synthesized from these 93 studies to answer the research questions and map the contribution in practice, as well as gaps and challenges.

### 2.1. Search strategy

Considering that this interdisciplinary question involves the fields of education, computer science, technology, and engineering, three major international databases were chosen: ACM Digital Library, IEEE Xplore, and Education Source. Since conference papers often have a higher publication impact than journals in computer science, both English-language journals and conference papers are included in the initial search process. The ACM Digital Library is the world's most comprehensive database of full-text articles in computing and information technology, making it ideal for the topic of this study. The IEEE Xplore Digital Library is the most common database of research literature in engineering and technology. Together the first two databases ensure that both computing and engineering perspectives on automation techniques for our topic are searchable. Education Source was selected as the world's largest full-text research database designed for education students, professionals, and policymakers. Additionally, the American Society for Engineering Education (ASEE) Annual Conference & Exposition was chosen as a supplementary database because of its leading position in the field of engineering education. While other databases such as Web of Science and Scopus have been widely used in systematic literature review, these databases have less literature on education or assessment, after repeated searching, and therefore we choose not to include them. Given the rapid advancement of AI techniques, particularly after the emergence of large-scale pre-trained language models like BERT (Bidirectional Encoder Representations from Transformers) (Devlin et al., 2018), researchers at Google have proposed in 2018 that the capability of machines to comprehend and learn from text has significantly improved. Accordingly, the search years were limited from 2017-2023 to ensure we focused on the latest techniques. For the ASEE Annual Conference & Exposition database, each conference from 2017-2022 was included. The 2022 ASEE Conference was the latest conference at the time of research. The initial search string, including five dimensions, was finalized and listed below in Table 1. The search was carried out on March 12, 2023.

**Table 1**
*Initial Search Strings*

| Topic | Search terms (Boolean operator: AND) |
|---|---|
| Application setting | automat* |
| Application setting | assess* OR evaluation OR grading OR scoring |
| Artificial intelligence | AI OR "artificial intelligence" OR NLP OR "natural language processing" |
| Application setting | "text-based response" OR "text-based question" OR "textual response" OR "text response" OR "short answer" OR "open ended question" OR "descriptive question" OR "constructed response" |
| Education level | "higher education" OR university OR college OR undergraduate OR "post-secondary" |

**2.2. Screening and filtering process**

The same search strings were applied in each of the four databases, and all records were kept and downloaded. Because the ASEE Annual Conference & Exposition database provides a relevance score for each paper, an additional criterion was applied to the ASEE database: only the records with a relevance score greater than 0.20 were selected for the review. The number of records from each database after searching is shown in Table 2. We found 365 studies from ACM Digital Library, 238 from IEEE Xplore, 212 from Education Source, and 23 from ASEE Annual Conference & Exposition.

Based on the topic of this systematic review, inclusion and exclusion criteria were developed by considering the research topic, application field, types of data, and relevance to the research questions of this study. Eight specific inclusion and exclusion criteria were used for screening and filtering the studies found from the databases, as shown in Table 3. Computer code and handwriting code were not considered for this review. Grading without the assistance of AI was not counted as an automatic assessment. In addition, non-empirical studies without research questions or primary data analysis were excluded.

**Table 2**
*Numbers of Papers by Database Searched*

| Databases | *n* |
|---|---|
| ACM Digital Library | 365 |
| IEEE Xplore | 238 |
| Education Source | 212 |
| ASEE Annual Conference & Exposition | 23 |

**Table 3**
*Inclusion and Exclusion Criteria*

| Inclusion/ Exclusion Criteria | Position |
|---|---|
| Criteria 1: No direct connection (not within the scope) | yes/ no/ not sure |
| Criteria 2: Does the study present a primary study of automatic assessment of student answers? (Human grading not included) | yes/ no/ not sure |
| Criteria 3: Is the assessment based on textual responses? (Computer code/handwriting text is not included) | yes/ no/ not sure |
| Criteria 4: Is the data sample adopted from students/teachers? | yes/ no/ not sure |
| Criteria 5: Can the study be applied in post-secondary education? | yes/ no/ not sure |
| Criteria 6: Does the main content answer at least one of the research questions? (Non-empirical papers are excluded) | yes/ no/ not sure |
| Criteria 7: Conference index/schedule/review articles are excluded (not a research paper) | yes/ no/ not sure |
| Criteria 8: Book/book sections are excluded | yes/ no/ not sure |

### 2.3. Data extraction process

All the records/citations found from the four databases were uploaded into the literature management software EndNote 20.3 to extract the target data after the initial search. A duplicate check was conducted for these 838 records, and 6 duplicates were removed. The titles and abstracts of these 832 articles were then screened for the first time based on the inclusion and exclusion criteria. We filtered out 695 studies (specific excluded reasons are clarified on the left side of Figure 2), and 137 studies were selected for relevance to this study (31 from ACM, 51 from IEEE, 48 from Education Source, and 7 from ASEE). Full texts of these 137 studies were then retrieved and re-screened on their full content based on the same inclusion and exclusion criteria. In the second screening process, we filtered out 44 studies (specific excluded reasons are clarified on the left side of Figure 2). Finally, 93 studies were included and synthesized in the systematic review. The complete PRISMA flowchart of the search process with inclusion/exclusion decisions is shown in Figure 2. The first author conducted the literature search. Three researchers met regularly to discuss the screening process and review screening decisions to ensure inclusion and exclusion criteria were followed. The researchers used an open coding approach to identify topics in key sections of each paper, develop codes for these topics, and group codes into themes.

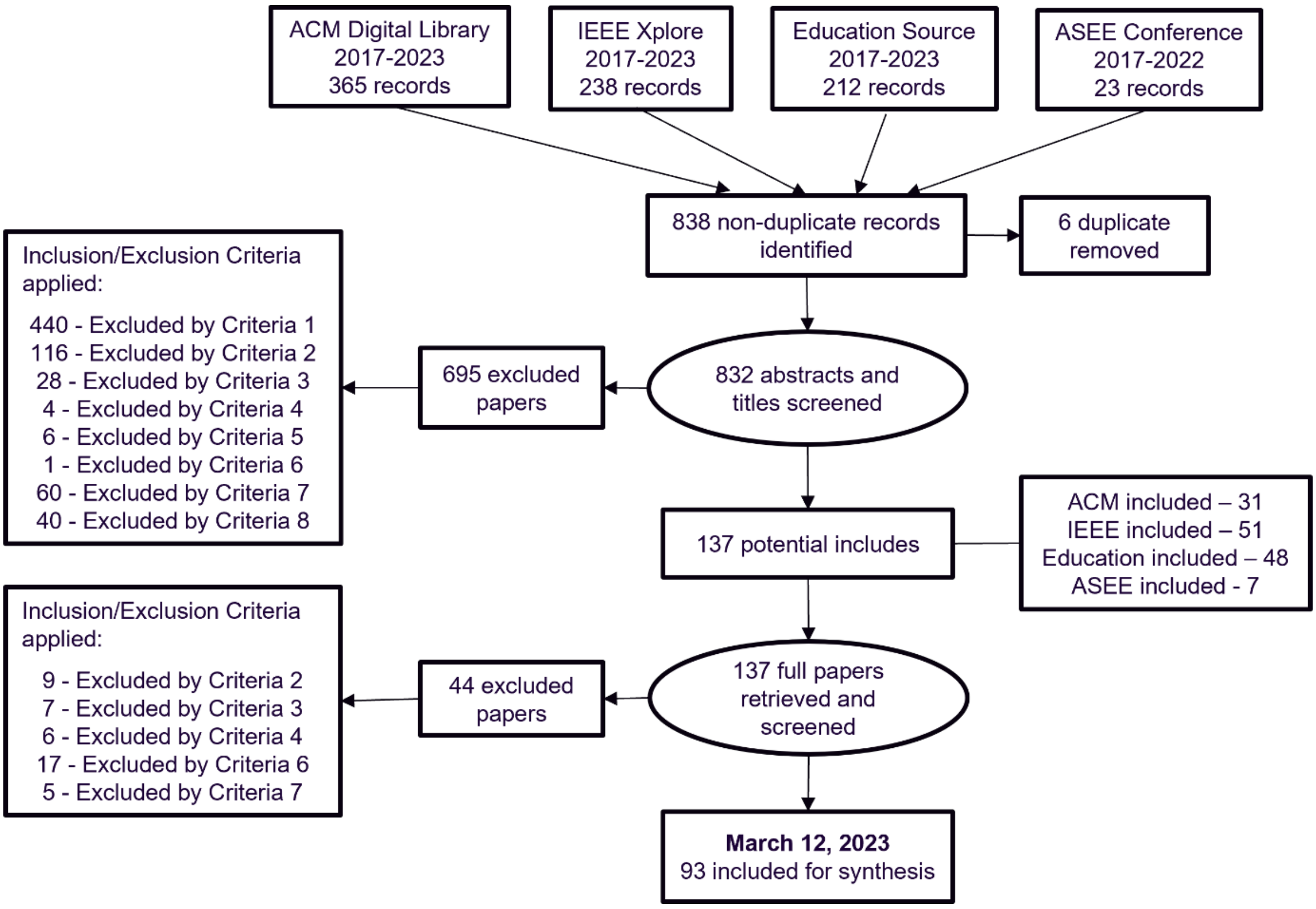

*Figure 2.* PRISMA flowchart of the search process and inclusion/exclusion decisions (see Section 2.2)

## 3. Results

### 3.1. RQ1: What types of automated assessment systems can be identified using input, output, and processing methods?

#### 3.1.1. Text-based automated assessment systems (TBAAS) types

A qualitative thematic analysis methodology based on the Input-Process-Output (IPO) framework, as mentioned in Section 1.1, was adopted to synthesize the various kinds of text-based automated assessment systems. These applications and tools can be seen as IPO systems. To analyze and integrate all studies, each study was parsed and simplified into three core sections: input, process, and output. In the PRISMA flow, input has already been restricted to student text information. Specifically, across all studies, we found that most input of the TBAAS were student responses to open-ended questions and can be categorized as short-answer, conceptual understanding, constructed responses, and writing text. The process of an automatic assessment system is the methodology of its text-based evaluation process. AI and NLP-based technology were applied to understand and analyze text information during this process. The outputs of a TBAAS were found to be classification labels, scores/grades, textual feedback, or further guidance. Several of these TBAAS include more than one type of output. The main feature of the output of TBAAS is whether the system provides textual feedback to the users or simply generates a score or the "right answer".

According to the IPO character of these systems, this iterative coding process led to the following five types of text-based automated assessment systems (TBAAS), as shown in Table 4: 1) Automatic Grading System; 2) Automatic Classifier; 3) Automated Feedback System; 4) Automated Writing Evaluation System; and 5) Multimodal Evaluation System. 1), 2), and 3) are categorized considering the characteristics of the outputs. The input to these systems is the students' short answers or conceptual questions. However, the outputs provided by the system focus on different facets; 4) is categorized by its specialized input text type, usually a student's written essay; and 5) is categorized by the systems designed for multi-input or providing multi-output. For each TBAAS category, the explanation and related studies are summarized in Table 4, and some of the selected highly-cited studies and best examples of each type of TBAAS are illustrated in the following sections. Studies are identified by their index number in Appendix B.

**Table 4**
*Text-based automated assessment system (TBAAS)*

| TBAAS type | Explanation | *n* | Studies |
|---|---|---|---|
| Automatic Grading System | Designed for evaluating student learning outcomes to grade student responses; the output is a grade/score (without textual feedback) | 39 | [1], [2], [3], [4], [5], [6], [11], [12], [19], [22], [25], [26], [27], [28], [29], [33], [34], [35], [38], [39], [43], [47], [50], [58], [60], [61], [62], [63], [65], [70], [71], [72], [75], [78], [82], [83], [84], [89], [91] |
| Automatic Classifier | Classifies student/teacher textual responses into different labels or categories (the label is not a score/grade) | 22 | [8], [10], [15], [17], [18], [36], [40], [41], [48], [51], [54], [64], [66], [74], [76], [77], [80], [81], [85], [87], [92], [93] |
| Automatic Feedback System | Designed for evaluating student learning outcomes, focusing on providing textual/visual feedback and guidance | 20 | [9], [16], [24], [30], [31], [32], [37], [42], [44], [45], [46], [52], [55], [56], [57], [67], [69], [73], [86], [90] |
| Automated Writing Evaluation System | Designed for automated essay evaluation to assess linguistic proficiency and improve students' quality of writing | 8 | [7], [20], [23], [49], [53], [59], [68], [88] |
| Multimodal Evaluation System | Designed for multi-input (e.g., drawing and writing input) or providing multi-output (e.g., both scores and visual feedback) | 4 | [13], [14], [21], [79] |

#### 3.1.1.1. Automatic grading system

The automatic grading system was the most common TBAAS found in our review. Dumal et al. (2017) proposed an e-learning system to automate essay question grading. The system has a Web-based answer generation module for predicting the best answers for short questions using NLP techniques. The system provides a numerical score after comparing keywords and semantic similarity between students' answers and the generated golden answers. The authors calculated the correlation coefficient between the actual score given by teachers and the AI-generated score as an indicator to evaluate the system's performance. The correlation coefficient was greater than 0.7. Erickson et al. (2020) focused on assessing students' responses/explanations for solving mathematics problems. Their TBAAS utilized ML and deep learning techniques to predict scores from student open responses. All models had an accuracy of at least 37% above chance in classifying and predicting the student's grade. Tulu et al. (2021) focused on grading short answer-based exams in a computer engineering course. They proposed an Automatic Short Answer Grading (ASAG) method by using the Manhattan LSTM network to calculate sentence similarity between student answers and correct answers. In particular, the authors used the SemSpace algorithm as a synset-based sense embedding approach to determining the sense vectors. Their TBAAS resulted in a 23% mean absolute error and 0.15 Pearson correlation, which was far from their expectation. Balaha and Saafan (2021) proposed an automatic exam correction framework for essays, multiple-choice questions, and equation matching. The best-achieved accuracy for short answer datasets in their experiments was 77.95% with the pre-trained large language model by the USE, and the proposed equations similarity checker algorithm (named "HMB-MMS-EMA") reported 100% accuracy on their proposed dataset (named "HMB-EMD-v1"). For most ASAG systems, detecting text similarity is important before predicting scores. Sahu and Bhowmick (2020) presented a comparative study of text similarity features such as semantic similarity, lexical overlap, alignment-based features, etc. and regression models toward ASAG based on the University of North Texas dataset.

#### 3.1.1.2. Automatic classifier

The automatic classifier was the second most common TBAAS. The system's input is usually the open-ended questions from questionnaires or tests; the output is different categorical labels. This type of TBAAS can help analysts automatically summarize the answers to the target themes, quickly find the common answers, and save time processing the questionnaires. Buenaño-Fernandez et al. (2020) proposed a methodology to collect valuable information from teacher self-assessment surveys based on topic modeling and text network modeling.

This system identified twelve topics describing the substantive content of unstructured textual survey responses. McDonald et al. (2020) used the text analysis software Quantext as a process to deal with students' free text comments on evaluations of teaching and courses, and categorized their responses into 4 labels which indicate students' judgment. Xing et al. (2020) classified student-written scientific arguments on the Albedo Effect, a challenging scientific phenomenon, using a text mining technique called Latent Dirichlet Allocation. Jescovitch et al. (2021) focused on classifying constructed responses according to 5 learning progression levels in an undergraduate STEM physiology context. They compared the performance between analytic and holistic coding approaches before making a classification rubric. Two specialists in physiology carefully divided the holistic rubric definitions into smaller conceptual pieces to create the analytic coding rubrics, which indicates better performance than a holistic rubric. Somers et al. (2021) used transformer-based NLP models to evaluate student conceptual understanding. The output of this TBAAS is their customized level of understanding with more than 90% accuracy. Furthermore, Zhu et al. (2022) used a pre-trained and fine-tuned BERT model to encode the answer text and achieve multiple classifications. The results show the proposed BERT-based model outperforms the baseline systems in their study regarding scoring accuracy. Yeruva et al. (2023) used a triplet loss-based Siamese network for classifying students' correct and incorrect answers.

#### 3.1.1.3. Automatic feedback system

The core of automated feedback systems is to provide textual feedback for students to revise or improve their responses. These responses can be conceptual understanding, constructed responses, scientific explanations, or revision suggestions. Krause et al. (2017) introduced a semi-automated method to help feedback providers to improve the quality of their feedback through self-assessment. The TBAAS extracts eight feature categories from their natural language model, including length, specificity, complexity, rarity, justification, active, subjective, and sentiment, to help with the critique-style guide. This guided intervention was helpful compared to the control group without guidance. Lee et al. (2019) adopted an NLP-enabled formative feedback system called HASbot to support student argumentation modifications in the science classroom. The HASbot feedback system output a machine score (diagnostic) and suggestive feedback toward student answers. Furthermore, Lee et al. (2021) studied simulation-based scientific argumentation tasks. As a design study, they investigated whether automated simulation feedback would be a beneficial addition to the existing automated argument feedback. The authors discussed important design lessons about utilizing ML to create automated scoring models and improving simulation feedback effectively in the classroom. Ruan et al. (2019) proposed a QuizBot system to help students learn factual knowledge in science, safety, and English vocabulary. This TBAAS consisted of a quiz mode and a casual chat mode, and the study showed that students who used QuizBot could recognize and remember almost 20% more accurate answers than those who used the flashcard app. In addition, Xia and Zilles (2023) proposed a novel approach to provide instantaneous feedback that helps the student improve their mathematical written statements by scaffolding the process of building a statement and restricting the number of choices students have compared to free text writing. They evaluated this tool in an undergraduate algorithms course and observed an improvement in students' mean test scores, which increased from 7.2/12 to 9.2/12 between the pre-test and the post-test.

#### 3.1.1.4. Automated writing evaluation system

Automated writing evaluation systems are characterized by their unique input type, typically a student's written essay, dedicated to assessing and enhancing students' English writing skills. *Criterion*® is the most widely recognized Automated Writing Evaluation program to assess students' English writing, and it is used by Educational Testing Service (ETS) to assess essays. The automated scoring engine of its program is *e-rater*®, uses NLP to extract four macro features (grammar, usage, mechanics, and style) from texts and then uses those features to automatically calculate essay scores (Chen et al., 2017; McCaffrey et al., 2022). Chien-Yuan (2017) investigated the effectiveness of an Initiation-Response-Feedback (IRF) based English grammar learning assistant system to enhance students' language learning and correct grammatical errors. The research found the IRF system significantly increased learning outcomes than the control group in an English grammar achievement test. Yannakoudakis et al. (2018) presented the creation and evaluation of an automated writing placement system for English language learners, leveraging supervised machine learning to predict an individual's general linguistic proficiency across the full Common European Framework of Reference scale by identifying highly predictive textual features. They then further integrated their model into W&I, a web-based program that automatically offers diagnostic feedback to non-native English language learners at various levels of granularity. It is openly accessible without charge to the public. In addition to English writing, Alqahtani and Alsaif (2019) focused on Arabic and proposed a rule-based TBAAS to evaluate Arabic language essays automatically. Their grading criteria include surface-based and text-processing standards such as spelling, punctuation, structure, coherence, and style. According to the data gathered by their system, 73% of the essay's overall scores were accurate.

##### 3.1.1.5. Multimodal evaluation system

For multi-input TBAAS, Chen et al. (2017) proposed an automatic evaluation system for university computer virtual experiment reports. The question types of this TBAAS were multiple-choice, fill-in-the-blank, and short-answer questions. For short answer questions, the output was a score of students' answers. Smith et al. (2019) proposed an integrated multimodal assessment framework for evaluating students' understanding of physical phenomena based on both students' writing and drawing. Specifically, a convolutional neural network-based model was adopted to assess student writing, and a topology-based model was used to assess student drawing. This study implies that multimodal assessment can be a useful strategy for implementing the new generation of assessment approaches to assess students' responses formulated in several modalities. For multi-output TBAAS, Becerra-Alonso et al. (2020) presented the development of an educational tool EduZinc for grading and providing class and individual student reports. This application allows for the simultaneous management of (a) the creation of individualized learning products and (b) automatic grading. It can also notify teachers and tutors when a student is lagging or when a student is excelling in the course. Beasley et al. (2021) incorporated sentiment analysis and aspect extraction into the peer-review process to address the challenge of accurately grading open-ended assignments in large-scale online courses. This system can output sentiment scores, visualization & associated files with marked sentimental words from a student's review.

## 3.2. RQ2: What are the educational focus and research motivations of studies with automated assessment systems?

### 3.2.1. Usage domains

We identified the usage domain of each paper as the discipline that student data were drawn from or the discipline in which the assessment system was applied (see Figure 3). STEM domains included engineering, design, mathematics, science, medicine, and technology. Humanities domains included Arabic, Chinese, English, Indonesian, and Japanese languages, education, social studies, and business. Other domains were multidisciplinary from incorporating student data from multiple fields or from using non-educational data from online forums or publicly available data for model testing. More than half of the total studies were from STEM domains (55%). Computer science studies made up nearly half of STEM papers (45%), with 15 science studies (29%) and 10 engineering studies (20%). Humanities domains were represented by approximately a third of studies (32%), with the English language (30%) and Arabic language (17%) studies most common. Multidisciplinary papers (9%) were a smaller but notable usage domain.

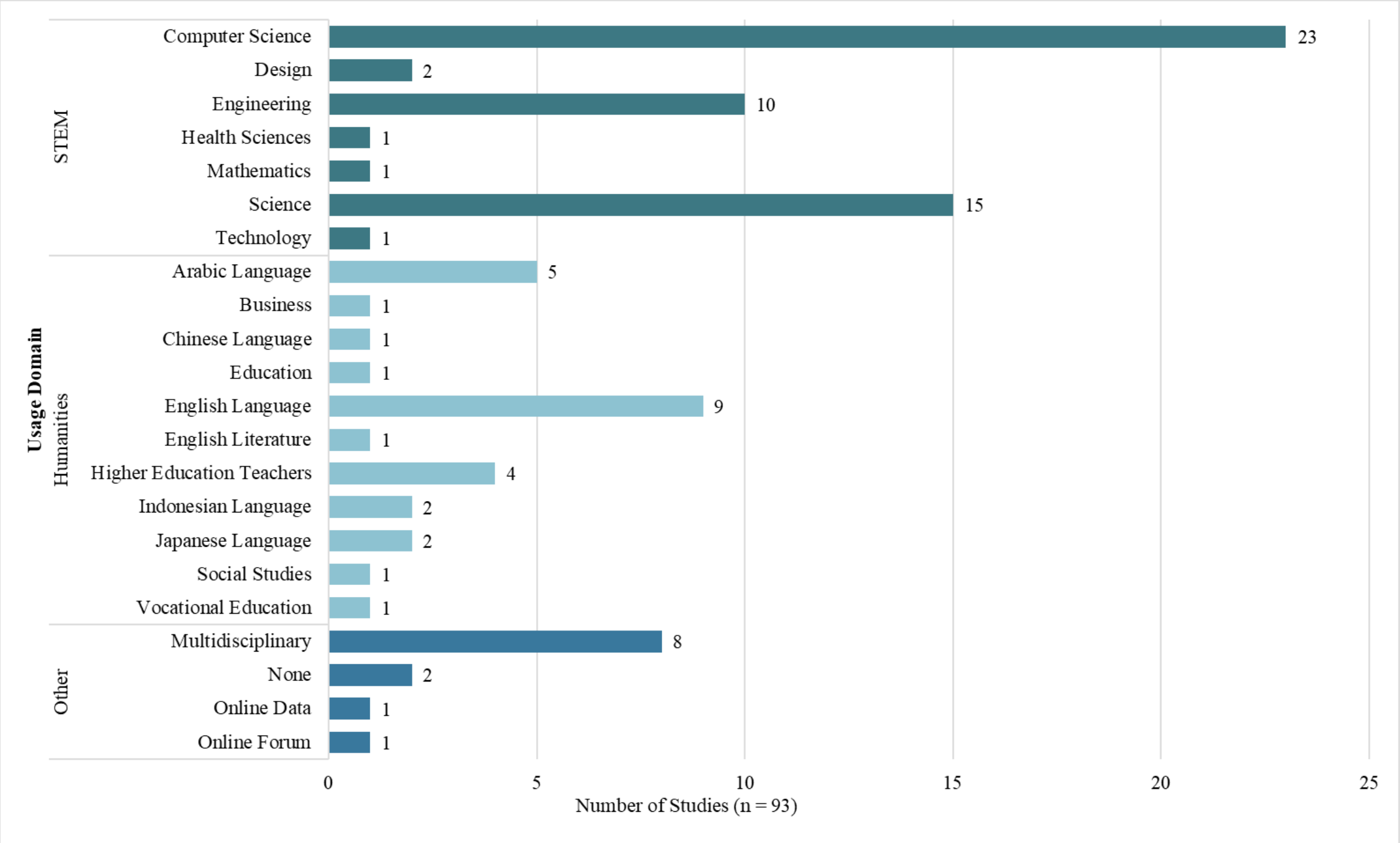

*Figure 3.* Usage domains of reviewed papers.

### 3.2.2. System features to support students

For the second research question, we investigated the learning needs and research motivations for designing automated text analysis systems in education. We determined how TBAAS addressed precise learning needs, the contributions of each paper in furthering education research knowledge, and how the systems were supported by learning concepts. Using an open coding approach and focusing on each paper's Introduction and Literature Review, we identified four broad learning needs and five specific research motivations. The findings of this research question are summarized in Table 5.

**Table 5**

*Learning Needs and Research Motivation for TBAAS*

| System Features | Categories | *n* | Studies |
|---|---|---|---|
| Learning Needs | Improve assessment, grading, and course evaluation | 31 | [3], [5], [8], [11], [13], [17], [21], [26], [27], [28], [33-35], [34], [35], [38], [39], [41], [43], [50], [53], [61], [68], [69], [70], [72], [74], [76], [77], [79], [89] |
| | Support specific content learning | 25 | [1], [6], [7], [10], [19], [30], [31], [40], [45], [47], [49], [57], [58], [59], [64], [65], [66], [67], [75], [80], [81], [84], [86], [87], [88] |
| | Reduce grading time, error, and effort | 19 | [2], [4], [12], [14], [18], [25], [29], [32], [36], [37], [62], [63], [71], [78], [82], [83], [85], [91], [92] |
| | Support students with personalization and feedback | 19 | [9], [15], [16], [20], [22], [23], [24], [42], [46], [48], [51], [52], [54], [55], [56], [60], [73], [90], [93] |
| | None | 1 | [44] |
| Research Motivation | Automate grading, review, and feedback | 27 | [1], [2], [3], [4], [6], [10], [12], [13], [14], [21], [27], [30], [32], [34], [38], [43], [51], [52], [57], [61], [62], [70], [72], [74], [78], [85], [92] |

| | | |
|---|---|---|
| Analyze meaning in open-ended text | 21 | [5], [7], [9], [11], [18], [19], [31], [36], [44], [47], [48], [56], [66], [69], [71], [73], [80], [87], [89], [91], [93] |
| Develop a system for learning | 17 | [15], [22], [23], [37], [39], [42], [45], [46], [58], [64], [67], [79], [81], [83], [84], [86], [88] |
| Validate a method or tool with human input | 15 | [16], [25], [28], [29], [33], [35], [53], [54], [55], [59], [63], [68], [75], [76], [77] |
| Test metrics of a method or tool | 13 | [8], [17], [20], [24], [26], [40], [41], [49], [50], [60], [65], [82], [90] |

#### 3.2.2.1. Learning needs addressed by automated assessment systems

We found that learning needs addressed by automated systems were most often centered around improving assessment practices, grading methods, and course evaluation (31%). These studies cited automatic scoring or response systems as a potential resource for better course assessment and grading practices. For example, Smith et al. (2019) proposed a multimodal assessment framework for assessing learning with science students' writing and drawings. Improving the quality of online or open education by interpreting text-based responses was another learning need within this category. Balaha and Saafan (2021) and Beasley et al. (2021) studied the need for automatic assessment in massive open online courses. A second group of papers focused on supporting specific content learning or analysis (27%). This topic addressed the need for improved assessment practices in a particular domain, such as Arabic, Chinese, English, Indonesian, and Japanese language learning, scientific argumentation, conceptual understanding, or design feedback. These studies connected automated assessment systems with the challenge of assessing learning in these domains. They explained why assessment is challenging due to scalability, complex subject material, or a lack of research with automated tools. Reducing the time, effort, and error for educators in grading text-based responses was a third justification for systems (20%). These systems were designed to take the burden of grading off teachers and increase the scalability of grading short answer and essay questions (Aini et al., 2018). For example, Galassi and Vittorini (2021) described the challenge of grading and providing detailed feedback for data science assignments. Studies have acknowledged instructors' challenges using complex analytics from learning management systems for instruction. Preventing error and bias by reducing subjectivity was another motivation for automated assessment. Hucko et al. (2018) proposed a method for real-time analysis of student responses during interactive lecture. A fourth group of learning needs was for improving the quality of student learning through personalization and feedback (20%). These studies focused on challenges such as analyzing student reflections, facilitating meaningful interactions in large courses, and identifying misconceptions. Studies recognized the need for guidance through formative assessment and engagement, with automated assessment systems as a potential solution to ensuring immediate and effective feedback.

#### 3.2.2.2. Research motivation for system studies

The largest number of studies were motivated by grading, scoring, and feedback challenges in automated assessment (29%). Various types of feedback included responding to student explanations of concepts (Auby et al., 2022), formative feedback for writing assignments (Hellman et al., 2019), and a formative assessment of scientific argumentation (Mao et al., 2018). These studies often proposed an improved approach to grading subjective, open-ended test questions. The second most significant area of studies was the motivation to research new methods of analyzing meaning in open-ended texts (23%). Texts were not as structured as test questions, but included feedback and evaluations. For example, Katz et al. (2021) were motivated to process large amounts of open-ended course feedback. A third type of study was motivated to develop complete learning systems with context and/or multiple indicators (18%). These systems had a long-term educational goal motivating the assessment system, such as developing writing skills, argumentation, or digital literacy. For example, Sung et al. (2021) developed a multimodal evaluation of students' conceptual learning with an augmented reality application. A fourth group of studies was often motivated to develop and validate a tool or method with human input (16%). These studies directly referenced human-developed grading systems and examined system performance relative to it (Hoblos, 2020). The smallest set of studies was motivated by testing and improving model features for reliability and agreement alone, rather than broader educational challenges (14%).

#### 3.2.2.3. Learning concepts informing system design

Learning concepts were coded using an open coding approach and focusing on the Literature Review across all papers. Because some studies were coded for multiple concepts, the total *n* is greater than 93, and we report the number of codes within each theme of learning concepts. Referenced learning themes by code frequency across all papers are shown in Figure 4. Feedback, discussion, and review was the theme with the most codes (28 codes). Example codes included dialogue for language learning, providing feedback to students, internal feedback on teamwork, design review, and peer grading. The second theme was from the educational measurement and testing fields, including psychometric testing (18 codes). These papers focused on developing and validating automated scoring and automated feedback systems from an educational assessment background. A third learning concept theme was critical thinking, problem-solving, scientific argumentation, and misconceptions (14 codes). This area informed systems that were meant to uncover student thinking and reasoning patterns in learning contexts. A fourth theme of learning concepts focused on matching instruction with student needs by studying engagement, cognitive load, and scaffolding (9 codes). Systems informed by these concepts were designed to directly support students by understanding individual learning. The fifth identified theme was on domain-specific instruction (9 codes), where systems supported assessment of a single topic such as language learning, digital literacy, or medicine. The last learning concept theme was on supporting student reflection (6 codes). This theme contained codes for systems which facilitate course evaluation and self-assessment as well as metacognition. Papers without learning concept backgrounds (33 codes) were primarily motivated by developing and testing new systems.

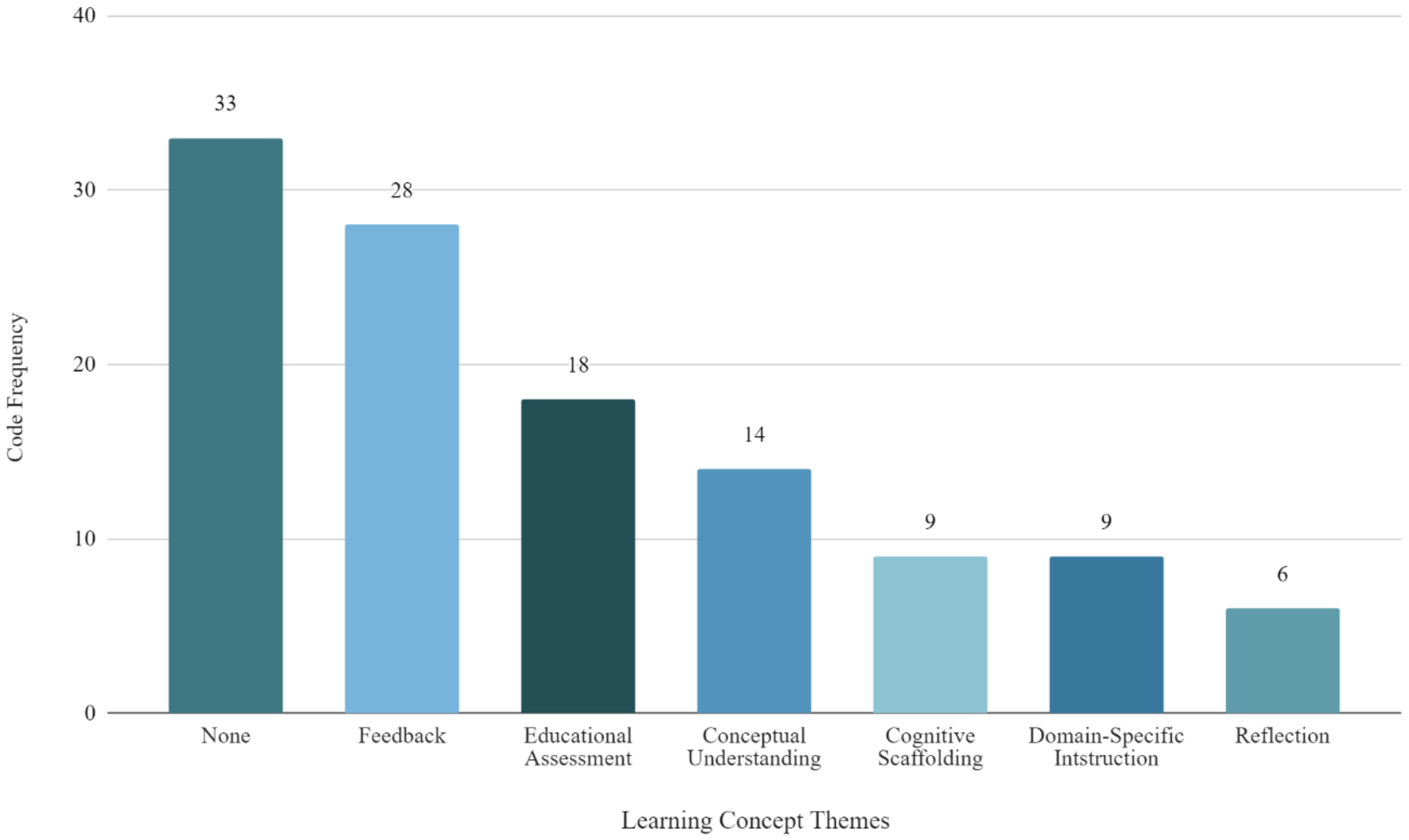

*Figure 4.* Theme frequencies of learning concepts referenced in collected papers.

### 3.3. RQ3: What are the reported research outcomes in automated assessment systems, and what are the next steps for educational application?

To explore the learning outcomes for automated assessment systems, we identified how each paper reported the key results of model performance. These findings are summarized as the main strengths of the systems in terms of what the authors achieved. Suggested future directions where these findings lead are also reported, as these represent ongoing work in education research integrating automated assessment into learning contexts. Using an open coding approach and focusing on the Results and Conclusions of each paper, we identified five major system performance outcomes and four proposed future directions across studies. Results are reported in Table 6.

**Table 6**

*Reported Model Performance and Future Education Directions*

| Outcomes | Categories | *n* | Studies |
|---|---|---|---|
| System Performance | Benchmark metrics | 48 | [1], [2], [4], [7], [9], [12], [13], [15], [16], [19-22], [25], [26], [28], [30], [34], [36], [39-42], [44], [49], [51], [52], [54-56], [61], [62], [65], [69-71], [75], [76], [80-82], [83], [87], [89-93] |
| | Multiple model comparison | 14 | [3], [6], [11], [14], [18], [27], [47], [48], [60], [63], [72], [77], [79], [85] |
| | Reflection, insights, usability, guidelines | 13 | [8], [10], [17], [24], [31], [32], [37], [46], [53], [59], [64], [66], [73] |
| | Machine-human comparison | 12 | [5], [29], [33], [35], [38], [43], [50], [58], [68], [74], [78], [84], [88] |
| | Learning impact | 6 | [23], [45], [57], [67], [86] |
| Future Directions | Improve model performance and generalizability | 47 | [1], [2], [5-7], [11], [12], [15], [16], [18], [20], [21], [26], [28], [29], [33], [35], [38], [39], [43], [47], [50], [51], [54], [55], [58], [60-65], [70-72], [74-79], [82], [83], [89], [90], [92], [93] |
| | Improve pedagogy for the system | 21 | [3], [9], [22], [30-32], [40-42], [44], [46], [48], [52], [53], [56], [57], [67-69], [73], [80] |
| | Support teachers | 12 | [8], [10], [13], [19], [24], [25], [34], [36], [59], [84], [87], [91] |
| | Enhance student learning | 11 | [4], [14], [17], [23], [37], [45], [49], [66], [81], [86], [88] |
| | None | 2 | [27], [85] |

### 3.3.1. System performance outcomes

Studies cited a variety of outcomes to report model performance. The largest group reported quantitative benchmark metrics (52%). These studies reported metrics such as accuracy, reliability, efficiency, root mean squared error, agreement levels, and correlations among performance scores as performance measures. Another type of benchmark was whether the overall system implementation goal was achieved, such as clustering student feedback, predicting student responses, meaningful topic modeling, or detecting semantic similarity. These broad results were qualified by details about effective aspects of the systems, and studies also shared improved or mixed model performance based on features. A second group of studies compared performance across models or the performance of a multi-model or hybrid system (16%). Studies reported the performance of combined systems compared to simple models, effects of systems using multiple approaches, or decisions selecting a best-performing model among multiple tests. A third group also relied on insights successfully obtained from texts, usability testing results, or broad recommendations and guidelines for discussing system performance (14%). Outcomes were defined by perceptions of test users or students, as well as the effectiveness of the system in supporting this type of conversation. A fourth group of studies reported outcomes by correlating or comparing machine scores with human ratings from experts or instructors (14%). These studies reported agreement between human and machine ratings, the similarity between the two, and described hybrid human-machine system performance. Finally, a fifth group of studies measured system performance as learning outcomes (5%). Observable impacts on student performance, such as improved pre-post test scores and improved recall, measured the effectiveness of the assessment system. For example, Chien-Yuan (2017) measured improved learning between a control groups and students who learned English grammar with automatic feedback and dialogue. Learning impact was also demonstrated as the sensitivity of the assessment system to detect pre-post change, support in scaffolding students on writing tasks, and positive impacts of interventions across learning contexts.

### 3.3.2. Future educational and research applications

The collected studies suggested many possible future directions for automated assessment systems. Approximately half of the studies (51%) offered specific ideas for improving model performance and generalizability, including optimizing efficiency and accuracy, testing the system in new contexts and formats with additional datasets, comparing to other methods, more comprehensive application to diverse samples, and comparing approaches to other new methods. A focus on model performance alone meant these studies did not provide discussion or applications on how their system supported student learning, but rather emphasized metrics as indicators for future work. However, several studies (23%) intended to apply findings towards pedagogical next steps and educational research through improved system design. These studies planned to adopt the system into existing courses or interventions, support better student-teacher interactions with the system, and improve the quality of the learning experience overall. In this way, the success of their system could continue to reach more audiences and applications. Studies also recognized teachers (13%) as a focus of future work through more detailed learning insights, easier teaching processes, and better training and decision-making. Finally, specific student aspects (12%) were goals for studies to continue improving assessment systems. Published studies that clearly mentioned students also emphasized the need to promote motivation, engagement, long-term learning, scaffolding, and complex reasoning with their systems.

## 4. Discussion

Within the STEM disciplines, representational competence is key for students to consider questions or inputs in one representation, such as a graph, a sketch, or an equation, convert it to a suitable symbolic representation for reasoning, and obtain conclusions for deciding how to represent conclusions in a manner that is most meaningful to the questioner (Sinatra & Pintrich, 2003). Planning and constructing explanations and examples, with explicit evaluation and description of the results in an active learning environment, can facilitate learning in a way that goes far beyond multiple-choice questions (Prevost et al., 2013). However, this is a significant challenge for automated assessment systems to support. **For RQ1**, “What types of automated assessment systems can be identified using input, output, and processing methods?”, we found that automatic grading systems that assigned numerical grades were most common, followed by automatic classification of student responses from labels and systems to provide feedback. Relatively fewer systems supported complex thinking and reasoning through feedback. While short-answer questions were a common focus of these systems, the output did not give more profound insight into student responses. In addition, while many studies referenced learning concepts of feedback and review, few described conceptual understanding or representational thinking.

**For RQ2**, “What are the educational focuses and research motivations of studies with automated assessment systems?”, we found that, while slightly more than half of the research in automated assessment has been carried out in STEM, computer science is the most represented discipline. Studies from outside STEM were most often in languages or with multidisciplinary data. TBAAS can support classroom instruction at many levels of student engagement. Instructors can use TBAAS to evaluate higher-level text-based questions, and automated assessment applications are needed in engineering education for grading conceptual reasoning answers to constructed response questions. For instructors to effectively teach using automated assessment systems, they must decide when and how to assign repetitive lower-level learning to grading systems and when to implement feedback systems for higher-level conceptual learning. TBAAS can support conceptual understanding by analyzing students' written response' content, providing instructors with classifications by meaning, and promoting student reflection and metacognition. Alternatively, systems can support problem identification and troubleshooting with detailed knowledge of student responses, grades, and automatic feedback. Instructors should incorporate TBAAS into classroom practice for holistic feedback and evaluations, rather than numerical evaluation alone. It is necessary to apply the appropriate automated assessment system to a learning context and to investigate its impact beyond model performance. In this way, education will be intentional about using new applications and understanding the effectiveness of tools.

**For RQ3**, “What are the reported research outcomes in automated assessment systems, and what are the next steps for educational application?”, we found that model performance metrics were emphasized over educational applications by half of the studies, but that the remaining studies planned improvements to the learning platforms, teacher support, and student learning outcomes. Many ML-based approaches are already being adopted for the broad category of student advising, such as identifying at-risk students and early warning notifications, such as Bertolini et al. (2021) who provide a comprehensive list of ML techniques for different aspects of student advising-related activities. These can be adapted to classroom instruction, and many authors cited here are already carrying out these efforts. The first step in this direction is automatically classifying questions and answers. For example, the work of Alammary (2021) is a step in identifying whether the questions asked are sufficiently challenging. The work of Goncher et al. (2016) on evaluating textual answers to concept inventory questions in addition to multiple choice questions by automated means indicates the potential to gain a much deeper insight into how students think about concepts. This has always been a challenge for large classes where instructors cannot grade the written answers and have preferred multiple-choice questions alone. Goncher et al. (2016) point out that "the dissimilarity between the software output and the human coder output suggests that it may be unlikely that an automated assessment of textual data will be able to assess students' explanations as accurately as that of a human" (p. 219). This is an opportunity for TBAAS to work alongside instructors and provide holistic feedback and evaluations. Automated systems do not need to supplant or replace the human assessor; instead, it is sufficient for the automated system to recognize the acceptable answers so that the human grader only deals with those that the automated system is unsure of or identifies as incorrect. This will free up more of the assessors' time to focus on addressing misconceptions.

## 5. Implications

This study contributes to knowledge on the design elements and educational implications of AI-based assessment systems. It summarizes the most frequently developed and researched system types with comparison of the input, process, and output. It also highlights the depth of each system of graded information and analysis of text-based responses. In addition, this study adds to the technical conversation by researching the educational implications of systems in terms of their theoretical foundations, the research motivations for automatic assessment systems, and the interpretation of performance metrics and future directions in the context of educational research.

## 6. Limitations

The study's results must be viewed in the light of some limitations. First, during the PRISMA of this systematic review, books and book sections were not considered because they are often summaries and discussions of existing work rather than presentations of new results. Excluding book chapters may have limited the scope of our findings on automated assessment tools and methods. Second, we only searched for studies written in English, meaning studies related to this topic but written in other languages were not considered. This is a limitation of excluding potentially relevant studies on language alone. Third, although we used a comprehensive approach for the systematic review, the standard limitations of the systematic review process apply. For example, we choose four databases, and other databases may yield more papers. Also, the selection bias is primarily reduced but cannot be entirely removed from such a review process.

## 7. Conclusions

In this systematic review, following the PRISMA (Preferred Reporting Items for Systematic Reviews and Meta-Analyses) process (Page et al., 2021; Selçuk, 2019), a scientific and reproducible literature search strategy is adopted to find comprehensive literature on the specific topic of automatic assessment of text-based responses in post-secondary education. Based on explicit inclusion and exclusion criteria, 838 journal/conference research papers related to this topic were screened, and finally, 93 studies were included in the review. All the studies were synthesized according to an improved IPO (input-process-output) framework, and these educational applications are categorized into five types of TBAAS (text-based automated assessment systems) according to their different characteristics of the input, output, and processing methodologies.

An **Automatic Grading System**, exemplified by Ye and Manoharan (2018), is designed to evaluate student learning outcomes by scoring their responses. The primary output of this type of TBAAS is a numerical grade or score, without providing textual feedback. Dumal et al. (2017) and Balaha and Saafan (2021)'s studies provided good system design frameworks and reference value for future research on Automatic Grading Systems. In contrast, an **Automatic Classifier**, such as Buenaño-Fernandez et al. (2020) and McDonald et al. (2020), focuses on categorizing text-based responses into different labels, which are not numerical scores or grades. A transformer model-based classification approach has been supported by Sayeed and Gupta's (2022) findings, highlighting its effectiveness in automatic assessment. An **Automatic Feedback System**, like Lee et al. (2019), is more inclined towards offering real-time textual and visual feedback, along with guidance to students. As Ruan et al. (2019) notes, this type of system greatly improves learner engagement and performance. Distinctly, an **Automated Writing Evaluation System**, exemplified by Yannakoudakis et al. (2018), specifically addresses assessing and improvements in of students' English writing skills. This type of system's efficacy in evaluating students' writing proficiency on a large scale has been corroborated by Rupp et al. (2019). Additionally, a **Multimodal Evaluation System** is distinguished by its capability to handle multiple inputs and provide diverse outputs, as explored by Smith et al. (2019) and Becerra-Alonso et al. (2020), providing a comprehensive assessment and support framework.

Educational applications are largely used in Computer Science, Science, and English language education domains. ML-based models and NLP-based models are widely adopted to devise TBAAS. Among them, Criterion® is the most mature commercial software developed by ETS to analyze textual data and evaluate students' English writing. While nearly a third of studies did not reference a learning concept, the ones which were represented most often included feedback and discussion, reliability and validity, and problem-solving and critical thinking. Studies reported model performance metrics more often than learning impact and largely focused on improving model performance. However, many studies were also interested in developing the pedagogy around TBAAS for integration into learning contexts and were interested in specific ways of supporting students and teachers.

Text-based tests motivate students to become deep learners more than multiple-choice tests. Compared to using single-best-answer questions in assessment, short-answer question formats have demonstrated higher degrees of reliability and validity, and items are perceived as more authentic (Sam et al., 2018). TBAAS systems have the potential to help students develop conceptual thinking through writing and feedback. Future work in automated text assessment for AIEd applications will benefit from greater alignment with learning concepts and research-based pedagogy to support integration with teaching. While text-based assessment is more challenging for educators, the abundance of AIEd applications in recent years allows us to look forward to a promising future of automatic assessment of student work.

**Acknowledgements**
Funding sources will be added later.

**Declaration of Competing Interest**

The authors declare that they have no known competing financial interests or personal relationships that could have appeared to influence the work reported in this paper.

**Statement on Open Data and Ethics**

The data supporting the findings of this study are available from the corresponding author upon reasonable request. Because this study is a systematic review, our manuscript does not report on or involve the use of any animal or human data and did not require the authors to obtain ethics approval and consent.

## References

Aini, Q., Julianto, A. E., & Purbohadi, D. (2018). *Development of a Scoring Application for Indonesian Language Essay Questions* Proceedings of the 2018 2nd International Conference on Education and E-Learning, Bali, Indonesia. https://doi.org/10.1145/3291078.3291099 https://dl.acm.org/doi/pdf/10.1145/3291078.3291099

Alammary, A. S. (2021). LOsMonitor: A machine learning tool for analyzing and monitoring cognitive levels of assessment questions. *IEEE Transactions on Learning Technologies*, *14*(5), 640-652.

Alqahtani, A., & Alsaif, A. (2019, 10-12 Dec. 2019). Automatic Evaluation for Arabic Essays: A Rule-Based System. 2019 IEEE International Symposium on Signal Processing and Information Technology (ISSPIT),

Alvero, A., Arthurs, N., Antonio, A. L., Domingue, B. W., Gebre-Medhin, B., Giebel, S., & Stevens, M. L. (2020). AI and holistic review: informing human reading in college admissions. Proceedings of the AAAI/ACM Conference on AI, Ethics, and Society,

Auby, H., Koretsky, M., Shivagunde, N., & Rumshisky, A. (2022). WIP: Using Machine Learning to Automate Coding of Student Explanations to Challenging Mechanics Concept Questions. 2022 ASEE Annual Conference & Exposition,

Bai, X., & Stede, M. (2022). A Survey of Current Machine Learning Approaches to Student Free-Text Evaluation for Intelligent Tutoring. *International Journal of Artificial Intelligence in Education*, 1-39.

Balaha, H. M., & Saafan, M. M. (2021). Automatic Exam Correction Framework (AECF) for the MCQs, Essays, and Equations Matching. *IEEE Access*, *9*, 32368-32389. https://doi.org/10.1109/ACCESS.2021.3060940

Beasley, Z. J., Piegl, L. A., & Rosen, P. (2021). Polarity in the Classroom: A Case Study Leveraging Peer Sentiment Toward Scalable Assessment. *IEEE Transactions on Learning Technologies*, *14*(4), 515-525. https://doi.org/10.1109/TLT.2021.3102184

Becerra-Alonso, D., Lopez-Cobo, I., Gómez-Rey, P., Fernández-Navarro, F., & Barbera, E. (2020). EduZinc: A tool for the creation and assessment of student learning activities in complex open, online and flexible learning environments [Article]. *Distance Education*, *41*(1), 86-105. https://doi.org/10.1080/01587919.2020.1724769

Bertolini, R., Finch, S. J., & Nehm, R. H. (2021). Testing the impact of novel assessment sources and machine learning methods on predictive outcome modeling in undergraduate biology. *Journal of Science Education and Technology*, *30*, 193-209.

Buenaño-Fernandez, D., González, M., Gil, D., & Luján-Mora, S. (2020). Text Mining of Open-Ended Questions in Self-Assessment of University Teachers: An LDA Topic Modeling Approach. *IEEE Access*, *8*, 35318-35330. https://doi.org/10.1109/ACCESS.2020.2974983

Caiza, J. C., & Del Alamo, J. M. (2013). Programming assignments automatic grading: review of tools and implementations. *INTED2013 Proceedings*, 5691-5700.

Cardella, M. E., & Tolbert, D. (2014). " Problem solving" in engineering: Research on students' engineering design practices and mathematical modeling practices. 2014 IEEE Frontiers in Education Conference (FIE) Proceedings,

Chen, J., Zhang, M., & Bejar, I. I. (2017). An Investigation of the e‐rater® Automated Scoring Engine's Grammar, Usage, Mechanics, and Style Microfeatures and Their Aggregation Model [Article]. *ETS Research Reports Series*, *2017*(1), 1-14. https://doi.org/10.1002/ets2.12131

Chen, L., Chen, P., & Lin, Z. (2020). Artificial intelligence in education: A review. *IEEE Access*, *8*, 75264-75278.

Chen, S.-M., & Bai, S.-M. (2010). Using data mining techniques to automatically construct concept maps for adaptive learning systems. *Expert Systems with Applications*, *37*(6), 4496-4503.

Chien-Yuan, S. (2017). Investigating the Effectiveness of an Interactive IRF-Based English Grammar Learning System [Article]. *International Journal of Emerging Technologies in Learning*, *12*(11), 63-82. https://doi.org/10.3991/ijet.v12i11.7036

Devlin, J., Chang, M.-W., Lee, K., & Toutanova, K. (2018). Bert: Pre-training of deep bidirectional transformers for language understanding. *arXiv preprint arXiv:1810.04805*.

Diefes‐Dux, H. A., Zawojewski, J. S., Hjalmarson, M. A., & Cardella, M. E. (2012). A framework for analyzing feedback in a formative assessment system for mathematical modeling problems. *Journal of Engineering Education*, *101*(2), 375-406.

Dumal, P. A. A., Shanika, W. K. D., Pathinayake, S. A. D., & Sandanayake, T. C. (2017, 6-8 Dec. 2017). Adaptive and automated online assessment evaluation system. 2017 11th International Conference on Software, Knowledge, Information Management and Applications (SKIMA),

Efendi, T., Lubis, F. F., Putri, A., Waskita, D., Sulistyaningtyas, T., Rosmansyah, Y., & Sembiring, J. (2022). A Bibliometrics-Based Systematic Review on Automated Essay Scoring in Education. 2022 International Conference on Information Technology Systems and Innovation (ICITSI),

Erickson, J. A., Botelho, A. F., McAteer, S., Varatharaj, A., & Heffernan, N. T. (2020). *The automated grading of student open responses in mathematics* Proceedings of the Tenth International Conference on Learning Analytics & Knowledge, Frankfurt, Germany. https://doi.org/10.1145/3375462.3375523 https://dl.acm.org/doi/pdf/10.1145/3375462.3375523

Feng, S., & Law, N. (2021). Mapping artificial intelligence in education research: A network‐based keyword analysis. *International Journal of Artificial Intelligence in Education*, *31*, 277-303.

Galassi, A., & Vittorini, P. (2021). *Automated Feedback to Students in Data Science Assignments: Improved Implementation and Results* CHItaly 2021: 14th Biannual Conference of the Italian SIGCHI Chapter, Bolzano, Italy. https://doi.org/10.1145/3464385.3464387 https://dl.acm.org/doi/pdf/10.1145/3464385.3464387

Geigle, C., Zhai, C., & Ferguson, D. C. (2016). An exploration of automated grading of complex assignments. Proceedings of the Third (2016) ACM Conference on Learning@ Scale,

Gikandi, J. W., Morrow, D., & Davis, N. E. (2011). Online formative assessment in higher education: A review of the literature. *Computers & Education*, *57*(4), 2333-2351. https://doi.org/https://doi.org/10.1016/j.compedu.2011.06.004

Goncher, A. M., Jayalath, D., & Boles, W. (2016). Insights into students' conceptual understanding using textual analysis: A case study in signal processing. *IEEE Transactions on Education*, *59*(3), 216-223.

Gunstone, R. F., & Northfield, J. (1994). Metacognition and learning to teach. *International Journal of Science Education*, *16*(5), 523-537.

Hellman, S., Rosenstein, M., Gorman, A., Murray, W., Becker, L., Baikadi, A., Budden, J., & Foltz, P. W. (2019). *Scaling Up Writing in the Curriculum: Batch Mode Active Learning for Automated Essay Scoring* Proceedings of the Sixth (2019) ACM Conference on Learning @ Scale, Chicago, IL, USA. https://doi.org/10.1145/3330430.3333629 https://dl.acm.org/doi/pdf/10.1145/3330430.3333629

Hoblos, J. (2020, 14-16 Dec. 2020). Experimenting with Latent Semantic Analysis and Latent Dirichlet Allocation on Automated Essay Grading. 2020 Seventh International Conference on Social Networks Analysis, Management and Security (SNAMS),

Huang, A. Y., Lu, O. H., & Yang, S. J. (2023). Effects of artificial Intelligence–Enabled personalized recommendations on learners' learning engagement, motivation, and outcomes in a flipped classroom. *Computers & Education*, *194*, 104684.

Hucko, M., Gaspar, P., Pikuliak, M., Triglianos, V., Pautasso, C., & Bielikova, M. (2018, 23-25 Aug. 2018). Short Texts Analysis for Teacher Assistance During Live Interactive Classroom Presentations. 2018 World Symposium on Digital Intelligence for Systems and Machines (DISA),

Hwang, G.-J., Xie, H., Wah, B. W., & Gašević, D. (2020). Vision, challenges, roles and research issues of Artificial Intelligence in Education. In (Vol. 1, pp. 100001): Elsevier.

Ilgen, D. R., Hollenbeck, J. R., Johnson, M., & Jundt, D. (2005). Teams in organizations: From input-process-output models to IMOI models. *Annu. Rev. Psychol.*, *56*, 517-543.

Jescovitch, L. N., Scott, E. E., Cerchiara, J. A., Merrill, J., Urban-Lurain, M., Doherty, J. H., & Haudek, K. C. (2021). Comparison of Machine Learning Performance Using Analytic and Holistic Coding Approaches Across Constructed Response Assessments Aligned to a Science Learning Progression [Article]. *Journal of Science Education & Technology*, *30*(2), 150-167. https://doi.org/10.1007/s10956-020-09858-0

Kabudi, T., Pappas, I., & Olsen, D. H. (2021). AI-enabled adaptive learning systems: A systematic mapping of the literature. *Computers and Education: Artificial Intelligence*, *2*, 100017.

Kasneci, E., Seßler, K., Küchemann, S., Bannert, M., Dementieva, D., Fischer, F., Gasser, U., Groh, G., Günnemann, S., & Hüllermeier, E. (2023). ChatGPT for good? On opportunities and challenges of large language models for education. *Learning and Individual Differences*, *103*, 102274.

Katz, A., Norris, M., Alsharif, A. M., Klopfer, M. D., Knight, D. B., & Grohs, J. R. (2021). Using Natural Language Processing to Facilitate Student Feedback Analysis. 2021 ASEE Virtual Annual Conference Content Access,

Kochmar, E., Vu, D. D., Belfer, R., Gupta, V., Serban, I. V., & Pineau, J. (2022). Automated data-driven generation of personalized pedagogical interventions in intelligent tutoring systems. *International Journal of Artificial Intelligence in Education*, *32*(2), 323-349.

Kohl, P. B., & Finkelstein, N. D. (2005). Student representational competence and self-assessment when solving physics problems. *Physical Review Special Topics-Physics Education Research*, *1*(1), 010104.

Krause, M., Garncarz, T., Song, J., Gerber, E. M., Bailey, B. P., & Dow, S. P. (2017). *Critique Style Guide: Improving Crowdsourced Design Feedback with a Natural Language Model* Proceedings of the 2017 CHI Conference on Human Factors in Computing Systems, Denver, Colorado, USA. https://doi.org/10.1145/3025453.3025883 https://dl.acm.org/doi/pdf/10.1145/3025453.3025883

Kung, T. H., Cheatham, M., Medenilla, A., Sillos, C., De Leon, L., Elepaño, C., Madriaga, M., Aggabao, R., Diaz-Candido, G., & Maningo, J. (2023). Performance of ChatGPT on USMLE: Potential for AI-assisted medical education using large language models. *PLoS digital health*, *2*(2), e0000198.

Langley, P. (2019). *An integrative framework for artificial intelligence* Proceedings of the AAAI Conference on Artificial Intelligence,

Lee, H.-S., Gweon, G.-H., Lord, T., Paessel, N., Pallant, A., & Pryputniewicz, S. (2021). Machine Learning-Enabled Automated Feedback: Supporting Students' Revision of Scientific Arguments Based on Data Drawn from Simulation [Article]. *Journal of Science Education & Technology*, *30*(2), 168-192. https://doi.org/10.1007/s10956-020-09889-7

Lee, H. S., Pallant, A., Pryputniewicz, S., Lord, T., Mulholland, M., & Liu, O. L. (2019). Automated text scoring and real‐time adjustable feedback: Supporting revision of scientific arguments involving uncertainty [Article]. *Science Education*, *103*(3), 590-622. https://doi.org/10.1002/sce.21504

Liu, O. L., Brew, C., Blackmore, J., Gerard, L., Madhok, J., & Linn, M. C. (2014). Automated scoring of constructed‐response science items: Prospects and obstacles. *Educational Measurement: Issues and Practice*, *33*(2), 19-28.

Luckin, R. (2017). Towards artificial intelligence-based assessment systems. *Nature Human Behaviour*, *1*(3), 0028.

Mao, L., Liu, O. L., Roohr, K., Belur, V., Mulholland, M., Lee, H.-S., & Pallant, A. (2018). Validation of Automated Scoring for a Formative Assessment that Employs Scientific Argumentation [Article]. *Educational Assessment*, *23*(2), 121-138. https://doi.org/10.1080/10627197.2018.1427570

McCaffrey, D. F., Casabianca, J. M., Ricker‐Pedley, K. L., Lawless, R. R., & Wendler, C. (2022). Best Practices for Constructed‐Response Scoring [Article]. *ETS Research Reports Series*, *2022*(1), 1-58. https://doi.org/10.1002/ets2.12358

McDonald, J., Moskal, A. C. M., Goodchild, A., Stein, S., & Terry, S. (2020). Advancing text-analysis to tap into the student voice: a proof-of-concept study [Article]. *Assessment & Evaluation in Higher Education*, *45*(1), 154-164. https://doi.org/10.1080/02602938.2019.1614524

Nunes, A., Cordeiro, C., Limpo, T., & Castro, S. L. (2022). Effectiveness of automated writing evaluation systems in school settings: A systematic review of studies from 2000 to 2020. *Journal of Computer Assisted Learning*, *38*(2), 599-620.

Ouyang, F., & Jiao, P. (2021). Artificial intelligence in education: The three paradigms. *Computers and Education: Artificial Intelligence*, *2*, 100020.

Page, M. J., McKenzie, J. E., Bossuyt, P. M., Boutron, I., Hoffmann, T. C., Mulrow, C. D., Shamseer, L., Tetzlaff, J. M., Akl, E. A., Brennan, S. E., Chou, R., Glanville, J., Grimshaw, J. M., Hróbjartsson, A., Lalu, M. M., Li, T., Loder, E. W., Mayo-Wilson, E., McDonald, S., . . . Moher, D. (2021). The PRISMA 2020 statement: An updated guideline for reporting systematic reviews. *International Journal of Surgery*, *88*, 105906. https://doi.org/https://doi.org/10.1016/j.ijsu.2021.105906

Prevost, L. B., Haudek, K. C., Henry, E. N., Berry, M. C., & Urban-Lurain, M. (2013). Automated text analysis facilitates using written formative assessments for just-in-time teaching in large enrollment courses. 2013 ASEE Annual Conference & Exposition,

Ramesh, D., & Sanampudi, S. K. (2022). An automated essay scoring systems: a systematic literature review. *Artificial Intelligence Review*, *55*(3), 2495-2527.

Roselli, R. J., & Brophy, S. P. (2006). Experiences with formative assessment in engineering classrooms. *Journal of Engineering Education*, *95*(4), 325-333.

Ruan, S., Jiang, L., Xu, J., Tham, B. J.-K., Qiu, Z., Zhu, Y., Murnane, E. L., Brunskill, E., & Landay, J. A. (2019). *QuizBot: A Dialogue-based Adaptive Learning System for Factual Knowledge* Proceedings of the 2019 CHI Conference on Human Factors in Computing Systems, Glasgow, Scotland Uk. https://doi.org/10.1145/3290605.3300587
https://dl.acm.org/doi/pdf/10.1145/3290605.3300587

Rupp, A. A., Casabianca, J. M., Krüger, M., Keller, S., & Köller, O. (2019). Automated Essay Scoring at Scale: A Case Study in Switzerland and Germany [Article]. *ETS Research Reports Series*, *2019*(1), 1-23. https://doi.org/10.1002/ets2.12249

Sahu, A., & Bhowmick, P. K. (2020). Feature Engineering and Ensemble-Based Approach for Improving Automatic Short-Answer Grading Performance. *IEEE Transactions on Learning Technologies*, *13*(1), 77-90. https://doi.org/10.1109/TLT.2019.2897997

Sallam, M. (2023). ChatGPT utility in healthcare education, research, and practice: systematic review on the promising perspectives and valid concerns. Healthcare,

Sam, A. H., Field, S. M., Collares, C. F., van der Vleuten, C. P., Wass, V. J., Melville, C., Harris, J., & Meeran, K. (2018). Very‐short‐answer questions: reliability, discrimination and acceptability. *Medical Education*, *52*(4), 447-455.

Sayeed, M. A., & Gupta, D. (2022, 14-16 Dec. 2022). Automate Descriptive Answer Grading using Reference based Models. 2022 OITS International Conference on Information Technology (OCIT),

Selçuk, A. A. (2019). A guide for systematic reviews: PRISMA. *Turkish archives of otorhinolaryngology*, *57*(1), 57.

Shepard, L. A. (2005). Formative assessment: Caveat emptor. ETS Invitational Conference, The Future of Assessment: Shaping Teaching and Learning, New York, NY,

Sinatra, G. M., & Pintrich, P. R. (2003). *Intentional conceptual change*. Routledge.

Smith, A., Leeman-Munk, S., Shelton, A., Mott, B., Wiebe, E., & Lester, J. (2019). A Multimodal Assessment Framework for Integrating Student Writing and Drawing in Elementary Science Learning. *IEEE Transactions on Learning Technologies*, *12*(1), 3-15. https://doi.org/10.1109/TLT.2018.2799871

Somers, R., Cunningham-Nelson, S., & Boles, W. (2021). Applying natural language processing to automatically assess student conceptual understanding from textual responses [Article]. *Australasian Journal of Educational Technology*, *37*(5), 98-115. https://doi.org/10.14742/ajet.7121

Sung, S. H., Li, C., Chen, G., Huang, X., Xie, C., Massicotte, J., & Shen, J. (2021). How Does Augmented Observation Facilitate Multimodal Representational Thinking? Applying Deep Learning to Decode Complex Student Construct [Article]. *Journal of Science Education & Technology*, *30*(2), 210-226. https://doi.org/10.1007/s10956-020-09856-2

Tarricone, P. (2011). *The taxonomy of metacognition*. Psychology Press.

Tulu, C. N., Ozkaya, O., & Orhan, U. (2021). Automatic Short Answer Grading With SemSpace Sense Vectors and MaLSTM. *IEEE Access*, *9*, 19270-19280. https://doi.org/10.1109/ACCESS.2021.3054346

Wang, J., & Brown, M. S. (2008). Automated essay scoring versus human scoring: A correlational study. *Contemporary Issues in Technology and Teacher Education*, *8*(4), 310-325.

Xia, J., & Zilles, C. (2023). *Using Context-Free Grammars to Scaffold and Automate Feedback in Precise Mathematical Writing* Proceedings of the 54th ACM Technical Symposium on Computer Science Education V. 1, Toronto ON, Canada. https://doi.org/10.1145/3545945.3569728
https://dl.acm.org/doi/pdf/10.1145/3545945.3569728

Xing, W., Lee, H.-S., & Shibani, A. (2020). Identifying patterns in students' scientific argumentation: content analysis through text mining using Latent Dirichlet Allocation [Article]. *Educational Technology Research & Development*, *68*(5), 2185-2214. https://doi.org/10.1007/s11423-020-09761-w

Yannakoudakis, H., Andersen, Ø. E., Geranpayeh, A., Briscoe, T., & Nicholls, D. (2018). Developing an automated writing placement system for ESL learners [Article]. *Applied Measurement in Education*, *31*(3), 251-267. https://doi.org/10.1080/08957347.2018.1464447

Ye, X., & Manoharan, S. (2018, 26-28 Sept. 2018). Machine Learning Techniques to Automate Scoring of Constructed-Response Type Assessments. 2018 28th EAEEIE Annual Conference (EAEEIE),

Yeruva, N., Venna, S., Indukuri, H., & Marreddy, M. (2023). *Triplet Loss based Siamese Networks for Automatic Short Answer Grading* Proceedings of the 14th Annual Meeting of the Forum for Information Retrieval Evaluation, Kolkata, India. https://doi.org/10.1145/3574318.3574337
https://dl.acm.org/doi/pdf/10.1145/3574318.3574337

Zawacki-Richter, O. (2019). Systematic review of research on artificial intelligence applications in higher education – where are the educators? *International Journal of Educational Technology in Higher Education*, 16-39.

Zhu, X., Wu, H., & Zhang, L. (2022). Automatic Short-Answer Grading via BERT-Based Deep Neural Networks. *IEEE Transactions on Learning Technologies*, *15*(3), 364-375. https://doi.org/10.1109/TLT.2022.3175537

**Appendix A: List of acronyms and full terms**

| **Acronym** | **Full Terms** |
|---|---|
| AI | Artificial Intelligence |
| AIEd | Artificial Intelligence in Education |
| ASAG | Automatic Short Answer Grading |
| BERT | Bidirectional Encoder Representation from a Transformer |
| IPO | Input-Process-Output |
| LLM | Large Language Model |
| ML | Machine Learning |
| NLP | Natural Language Processing |
| PRISMA | Preferred Reporting Items for Systematic Reviews and Meta-Analysis |
| TBAAS | Text-Based Automated Assessment System |

**Appendix B: List of systematically reviewed papers**

| | Reference | Domain |
|---|---|---|
| 1 | Abdeljaber, H. A. (2021). Automatic Arabic short answers scoring using longest common subsequence and Arabic WordNet. *IEEE Access, 9, 76433-76445.* https://doi.org/10.1109/ACCESS.2021.3082408 | Arabic Language |
| 2 | Agarwal, M., Kalia, R., Bahel, V., & Thomas, A. (2021, 24-26 Sept. 2021). AutoEval: A NLP Approach for Automatic Test Evaluation System. *2021 IEEE 4th International Conference on Computing, Power and Communication Technologies (GUCON), Kuala Lumpur, Malaysia*, 1-6. https://doi.org/10.1109/GUCON50781.2021.9573769 | None |
| 3 | Ahmed, A., Joorabchi, A., & Hayes, M. J. (2022, 9-10 June 2022). On the Application of Sentence Transformers to Automatic Short Answer Grading in Blended Assessment. *2022 33rd Irish Signals and Systems Conference (ISSC*), Cork, Ireland, 1-6. https://doi.org/10.1109/ISSC55427.2022.9826194 | Computer Science |
| 4 | Aini, Q., Julianto, A. E., & Purbohadi, D. (2018). Development of a Scoring Application for Indonesian Language Essay Questions. *Proceedings of the 2018 2nd International Conference on Education and E-Learning, Bali, Indonesia*. https://doi.org/10.1145/3291078.3291099 | Indonesian Language |
| 5 | Alobed, M., Altrad, A. M. M., & Bakar, Z. B. A. (2021, 15-17 June 2021). An Adaptive Automated Arabic Essay Scoring Model Using the Semantic of Arabic WordNet. *2021 2nd International Conference on Smart Computing and Electronic Enterprise (ICSCEE), Cameron Highlands, Malaysia*, 45-54. https://doi.org/ICSCEE50312.2021.9498191 | Arabic Language |
| 6 | Alobed, M., Altrad, A. M. M., & Bakar, Z. B. A. (2021, 15-16 June 2021). A Comparative Analysis of Euclidean, Jaccard and Cosine Similarity Measure and Arabic Wordnet for Automated Arabic Essay Scoring. *2021 Fifth International Conference on Information Retrieval and Knowledge Management (CAMP), Kuala Lumpur, Malaysia*, 70-74. https://doi.org/10.1109/CAMP51653.2021.9498119 | Arabic Language |
| 7 | Alqahtani, A., & Alsaif, A. (2019, 10-12 Dec. 2019). Automatic Evaluation for Arabic Essays: A Rule-Based System. *2019 IEEE International Symposium on Signal Processing and Information Technology (ISSPIT), Ajman, United Arab Emirates*, 1-7. https://doi.org/10.1109/ISSPIT47144.2019.9001802 | Arabic Language |
| 8 | Alsharif, A., Katz, A., Knight, D., & Alatwah, S. (2022). Using Sentiment Analysis to Evaluate First-year Engineering Students Teamwork Textual Feedback. 2022 ASEE Annual Conference & Exposition. https://peer.asee.org/41460 | Engineering |
| 9 | Altoe, F., & Joyner, D. (2019, 23-25 Oct. 2019). Annotation-free Automatic Examination Essay Feedback Generation. *2019 IEEE Learning With MOOCS (LWMOOCS), Milwaukee, WI, USA,* 110-115. https://doi.org/10.1109/LWMOOCS47620.2019.8939630 | Computer Science |
| 10 | Auby, H., Koretsky, M., Shivagunde, N., & Rumshisky, A. (2022). WIP: Using Machine Learning to Automate Coding of Student Explanations to Challenging Mechanics Concept Questions. 2022 ASEE Annual Conference & Exposition. https://peer.asee.org/40507 | Engineering |
| 11 | Balaha, H. M., & Saafan, M. M. (2021). Automatic exam correction framework (AECF) for the MCQs, essays, and equations matching. *IEEE Access, 9*, 32368-32389. https://doi.org/10.1109/ACCESS.2021.3060940 | Computer Science |
| 12 | Bashir, M. F., Arshad, H., Javed, A. R., Kryvinska, N., & Band, S. S. (2021). Subjective answers evaluation using machine learning and natural language processing. *IEEE Access, 9*, 158972-158983. https://doi.org/10.1109/ACCESS.2021.3130902 | Computer Science |
| 13 | Beasley, Z. J., Piegl, L. A., & Rosen, P. (2021). Polarity in the classroom: A case study leveraging peer sentiment toward scalable assessment. *IEEE Transactions on Learning Technologies, 14*(4), 515-525. https://doi.org/10.1109/TLT.2021.3102184 | Computer Science |
| 14 | Becerra-Alonso, D., Lopez-Cobo, I., Gómez-Rey, P., Fernández-Navarro, F., & Barbera, E. (2020). EduZinc: A tool for the creation and assessment of student learning activities in complex open, online and flexible learning environments. *Distance Education, 41*(1), 86-105. https://doi.org/10.1080/01587919.2020.1724769 | Engineering |
| 15 | Becker, J. P., Kahanda, I., & Kazi, N. H. (2021). WIP: Detection of student misconceptions of electrical circuit concepts in a short answer question using NLP. 2021 ASEE Virtual Annual Conference Content Access. https://peer.asee.org/38076 | Engineering |
| 16 | Bernius, J. P., Krusche, S., & Bruegge, B. (2021). A Machine Learning Approach for Suggesting Feedback in Textual Exercises in Large Courses. *Proceedings of the Eighth ACM Conference on Learning @ Scale, Virtual Event, Germany*. https://doi.org/10.1145/3430895.3460135 | Engineering |
| 17 | Bhaduri, S., Soledad, M., Roy, T., Murzi, H., & Knott, T. (2021). A Semester Like No Other: Use of Natural Language Processing for Novice-Led Analysis on End-of-Semester Responses on Students' Experience of Changing Learning Environments Due to COVID-19. 2021 ASEE Virtual Annual Conference Content Access. https://peer.asee.org/36609 | Engineering |
| 18 | Buenaño-Fernandez, D., González, M., Gil, D., & Luján-Mora, S. (2020). Text mining of open-ended questions in self-assessment of university teachers: An LDA topic modeling approach. *IEEE Access, 8*, 35318-35330. https://doi.org/10.1109/ACCESS.2020.2974983 | Higher Education Teachers |
| 19 | Chaudhuri, N. B., Dhar, D., & Yammiyavar, P. G. (2022). A computational model for subjective evaluation of novelty in descriptive aptitude. *International Journal of Technology & Design Education, 32*(2), 1121-1158. https://doi.org/10.1007/s10798-020-09638-2 | Design |
| 20 | Chen, J., Zhang, M., & Bejar, I. I. (2017). An investigation of the e-rater® automated scoring engine's grammar, usage, mechanics, and style microfeatures and their aggregation model. *ETS Research Reports Series, 2017*(1), 1-14. https://doi.org/10.1002/ets2.12131 | English Language |
| 21 | Chen, Y., Liu, X., Huo, P., Li, L., & Li, F. (2017, 22-25 Aug. 2017). The design and implementation for automatic evaluation system of virtual experiment report. *2017 12th International Conference on Computer Science and Education (ICCSE), Houston, TX, USA*, 717-721. https://doi.org/10.1109/ICCSE.2017.8085587 | Science |
| 22 | Cheong, M. L. F., Chen, J. Y. C., & Dai, B. T. (2018, 4-7 Dec. 2018). Integrated Telegram and Web-based Forum with Automatic Assessment of Questions and Answers for Collaborative Learning. *2018 IEEE International Conference on Teaching, Assessment, and Learning for Engineering (TALE), Wollongong, NSW, Australia*, 9-16. https://doi.org/10.1109/TALE.2018.8615137 | Online Forum |
| 23 | Chien-Yuan, S. (2017). Investigating the effectiveness of an interactive IRF-based English grammar learning system. *International Journal of Emerging Technologies in Learning, 12(*11), 63-82. https://doi.org/10.3991/ijet.v12i11.7036 | English Language |
| 24 | Cunningham-Nelson, S., Laundon, M., & Cathcart, A. (2021). Beyond satisfaction scores: Visualising student comments for whole-of-course evaluation. *Assessment & Evaluation in Higher Education, 46(*5), 685-700. https://doi.org/10.1080/02602938.2020.1805409 | Health Sciences |
| 25 | Das, I., Sharma, B., Rautaray, S. S., & Pandey, M. (2019, 17-19 July 2019). An Examination System Automation Using Natural Language Processing. *2019 International Conference on Communication and Electronics Systems (ICCES), Coimbatore, India,* 1064-1069. https://doi.org/10.1109/ICCES45898.2019.9002048 | Computer Science |
| 26 | Devi, P. S., Sarkar, S., Singh, T. S., Sharma, L. D., Pankaj, C., & Singh, K. R. (2022, 8-10 July 2022). An Approach to Evaluating Subjective Answers using BERT model. *2022 IEEE International Conference on Electronics, Computing and Communication Technologies (CONECCT), Bangalore, India,* 1-4. https://doi.org/10.1109/CONECCT55679.2022.9865706 | Multidisciplinary |
| 27 | Dumal, P. A. A., Shanika, W. K. D., Pathinayake, S. A. D., & Sandanayake, T. C. (2017, 6-8 Dec. 2017). Adaptive and automated online assessment evaluation system. *2017 11th International Conference on Software, Knowledge, Information Management and Applications (SKIMA), Malabe, Sri Lanka*, 1-8. https://doi.org/10.1109/SKIMA.2017.8294135 | Online Data |
| 28 | Erickson, J. A., Botelho, A. F., McAteer, S., Varatharaj, A., & Heffernan, N. T. (2020). The automated grading of student open responses in mathematics. *Proceedings of the Tenth International Conference on Learning Analytics & Knowledge, Frankfurt, Germany*. https://doi.org/10.1145/3375462.3375523 | Mathematics |
| 29 | Forsyth, S., & Mavridis, N. (2021, 14-17 March 2021). Short Answer Marking Agent for GCSE Computer Science. *2021 IEEE World Conference on Engineering Education (EDUNINE), Guatemala City, Guatemala*, 1-6. https://doi.org/10.1109/EDUNINE51952.2021.9429163 | Computer Science |
| 30 | Fowler, M., Chen, B., Azad, S., West, M., & Zilles, C. (2021). Autograding "Explain in Plain English" questions using NLP. *Proceedings of the 52nd ACM Technical Symposium on Computer Science Education, Virtual Event, USA*. https://doi.org/10.1145/3408877.3432539 | Computer Science |
| 31 | Fwa, H. L. (2022). Enhancing Project Based Learning with Unsupervised Learning of Project Reflections. *Proceedings of the 5th International Conference on Digital Technology in Education, Busan, Republic of Korea*. https://doi.org/10.1145/3488466.3488480 | Computer Science |
| 32 | Galassi, A., & Vittorini, P. (2021). Automated Feedback to Students in Data Science Assignments: Improved Implementation and Results. *CHItaly 2021: 14th Biannual Conference of the Italian SIGCHI Chapter, Bolzano, Italy*. https://doi.org/10.1145/3464385.3464387 | Computer Science |
| 33 | Han, Y., Wu, W., Liang, Y., & Zhang, L. (2022). Peer Grading Eliciting Truthfulness Based on Auto-grader. *IEEE Transactions on Learning Technologies,* 1-12. https://doi.org/10.1109/TLT.2022.3216946 | Computer Science |
| 34 | Hellman, S., Rosenstein, M., Gorman, A., Murray, W., Becker, L., Baikadi, A., Budden, J., & Foltz, P. W. (2019). Scaling Up Writing in the Curriculum: Batch Mode Active Learning for Automated Essay Scoring. *Proceedings of the Sixth (2019) ACM Conference on Learning @ Scale, Chicago, IL, USA*. https://doi.org/10.1145/3330430.3333629 | Multidisciplinary |
| 35 | Hoblos, J. (2020, 14-16 Dec. 2020). Experimenting with Latent Semantic Analysis and Latent Dirichlet Allocation on Automated Essay Grading. *2020 Seventh International Conference on Social Networks Analysis, Management and Security (SNAMS), Paris, France*, 1-7. https://doi.org/10.1109/SNAMS52053.2020.9336533 | Computer Science |
| 36 | Hucko, M., Gaspar, P., Pikuliak, M., Triglianos, V., Pautasso, C., & Bielikova, M. (2018, 23-25 Aug. 2018). Short Texts Analysis for Teacher Assistance During Live Interactive Classroom Presentations. *2018 World Symposium on Digital Intelligence for Systems and Machines (DISA), Košice, Slovakia*, 239-244. https://doi.org/10.1109/DISA.2018.8490610 | Computer Science |
| 37 | Hurtig, N., Hollingsworth, J., & Scrivner, O. (2022, 28-31 March 2022). Visualization of Students' Solutions as a Sequential Network. *2022 IEEE Global Engineering Education Conference (EDUCON), Tunis, Tunisia*, 1189-1194. https://doi.org/10.1109/EDUCON52537.2022.9766502 | Higher Education Teachers |
| 38 | Ishioka, T., & Kameda, M. (2017). Overwritable automated Japanese short-answer scoring and support system. *Proceedings of the International Conference on Web Intelligence, Leipzig, Germany*. https://doi.org/10.1145/3106426.3106513 | Japanese Language |
| 39 | Jayawardena, R. R. A. M. P., Thiwanthi, G. A. D., Suriyaarachchi, P. S., Withana, K. I., & Jayawardena, C. (2018, 21-22 Dec. 2018). Automated Exam Paper Marking System for Structured Questions and Block Diagrams. *2018 IEEE International Conference on Information and Automation for Sustainability (ICIAfS), Colombo, Sri Lanka,* 1-5. https://doi.org/10.1109/ICIAFS.2018.8913351 | Computer Science |
| 40 | Jescovitch, L. N., Scott, E. E., Cerchiara, J. A., Merrill, J., Urban-Lurain, M., Doherty, J. H., & Haudek, K. C. (2021). Comparison of machine learning performance using analytic and holistic coding approaches across constructed response assessments aligned to a science learning progression. *Journal of Science Education & Technology, 30*(2), 150-167. https://doi.org/10.1007/s10956-020-09858-0 | Science |
| 41 | Katz, A., Norris, M., Alsharif, A. M., Klopfer, M. D., Knight, D. B., & Grohs, J. R. (2021). Using Natural Language Processing to Facilitate Student Feedback Analysis. 2021 ASEE Virtual Annual Conference Content Access. https://peer.asee.org/37994 | Engineering |
| 42 | Kolchinski, Y. A., Ruan, S., Schwartz, D., & Brunskill, E. (2018). Adaptive natural-language targeting for student feedback. *Proceedings of the Fifth Annual ACM Conference on Learning at Scale, London, United Kingdom*. https://doi.org/10.1145/3231644.3231684 | Science |
| 43 | Kosh, A. E., Greene, J. A., Murphy, P. K., Burdick, H., Firetto, C. M., & Elmore, J. (2018). Automated scoring of students' small-group discussions to assess reading ability. *Educational Measurement: Issues & Practice, 37*(2), 20-34. https://doi.org/10.1111/emip.12174 | English Language |
| 44 | Krause, M., Garncarz, T., Song, J., Gerber, E. M., Bailey, B. P., & Dow, S. P. (2017). Critique Style Guide: Improving Crowdsourced Design Feedback with a Natural Language Model. *Proceedings of the 2017 CHI Conference on Human Factors in Computing Systems,* Denver, Colorado, USA. https://doi.org/10.1145/3025453.3025883 | Design |
| 45 | Lee, H. S., Pallant, A., Pryputniewicz, S., Lord, T., Mulholland, M., & Liu, O. L. (2019). Automated text scoring and real-time adjustable feedback: Supporting revision of scientific | Science |
| | arguments involving uncertainty. *Science Education, 103*(3), 590-622. https://doi.org/10.1002/sce.21504 | |
| 46 | Lee, H.-S., Gweon, G.-H., Lord, T., Paessel, N., Pallant, A., & Pryputniewicz, S. (2021). Machine learning-enabled automated feedback: Supporting students' revision of scientific arguments based on data drawn from simulation. *Journal of Science Education & Technology, 30*(2), 168-192. https://doi.org/10.1007/s10956-020-09889-7 | Science |
| 47 | Leila, O., & Djamal, B. (2018, 28-30 Nov. 2018). A vector space based approach for short answer grading system. *2018 International Arab Conference on Information Technology (ACIT), Werdanye, Lebanon*, 1-9. https://doi.org/10.1109/ACIT.2018.8672717 | Arabic Language |
| 48 | Lo, S. L., Tan, K. W., & Ouh, E. L. (2021). Automated doubt identification from informal reflections through hybrid sentic patterns and machine learning approach. *Research & Practice in Technology Enhanced Learning, 16*(1), 1-24. https://doi.org/10.1186/s41039-021-00149-9 | Computer Science |
| 49 | Lopez, A. A., Guzman-Orth, D., Zapata-Rivera, D., Forsyth, C. M., & Luce, C. (2021). Examining the accuracy of a conversation-based assessment in interpreting English learners' written responses. *ETS Research Reports Series, 2021*(1), 1-15. https://doi.org/10.1002/ets2.12315 | English Language |
| 50 | Luchoomun, T., Chumroo, M., & Ramnarain-Seetohul, V. (2019, 8-11 April 2019). A Knowledge Based System for Automated Assessment of Short Structured Questions. *2019 IEEE Global Engineering Education Conference (EDUCON), Dubai, United Arab Emirates*, 1349-1352. https://doi.org/10.1109/EDUCON.2019.8725139 | Computer Science |
| 51 | Lv, G., Song, W., Cheng, M., & Liu, L. (2021, 18-20 June 2021). Exploring the Effectiveness of Question for Neural Short Answer Scoring System. *2021 IEEE 11th International Conference on Electronics Information and Emergency Communication (ICEIEC), Beijing, China,* 1-4. https://doi.org/10.1109/ICEIEC51955.2021.9463814 | Science |
| 52 | Mao, L., Liu, O. L., Roohr, K., Belur, V., Mulholland, M., Lee, H.-S., & Pallant, A. (2018). Validation of automated scoring for a formative assessment that employs scientific argumentation. *Educational Assessment, 23*(2), 121-138. https://doi.org/10.1080/10627197.2018.1427570 | Science |
| 53 | McCaffrey, D. F., Casabianca, J. M., Ricker-Pedley, K. L., Lawless, R. R., & Wendler, C. (2022). Best practices for constructed-response scoring. *ETS Research Reports Series, 2022*(1), 1-58. https://doi.org/10.1002/ets2.12358 | Other |
| 54 | McDonald, J., Moskal, A. C. M., Goodchild, A., Stein, S., & Terry, S. (2020). Advancing text-analysis to tap into the student voice: A proof-of-concept study. *Assessment & Evaluation in Higher Education, 45*(1), 154-164. https://doi.org/10.1080/02602938.2019.1614524 | Technology |
| 55 | Michalenko, J. J., Lan, A. S., & Baraniuk, R. G. (2017). Data-Mining Textual Responses to Uncover Misconception Patterns. *Proceedings of the Fourth (2017) ACM Conference on Learning @ Scale,* Cambridge, Massachusetts, USA. https://doi.org/10.1145/3051457.3053996 | Science |
| 56 | Nyomi, X., & Moccozet, L. (2022, 7-9 Nov. 2022). Anatomy of a large-scale real-time peer evaluation system. *2022 20th International Conference on Information Technology Based Higher Education and Training (ITHET), Antalya, Turkey*, 1-9. https://doi.org/10.1109/ITHET56107.2022.10032005 | Higher Education Teachers |
| 57 | Olivera-Aguilar, M., Lee, H. S., Pallant, A., Belur, V., Mulholland, M., & Liu, Ou L. (2022). Comparing the effect of contextualized versus generic automated feedback on students' scientific argumentation. *ETS Research Reports Series, 2022*(1), 1-14. https://doi.org/10.1002/ets2.12344 | Science |
| 58 | Poce, A., Medio, C. d., Amenduni, F., & Re, M. R. (2019, 26-27 Sept. 2019). Critical Thinking assessment: a first approach to the automatic evaluation. *2019 18th International Conference on Information Technology Based Higher Education and Training (ITHET), Magdeburg, Germany*, 1-8. https://doi.org/10.1109/ITHET46829.2019.8937353 | Multidisciplinary |
| 59 | Potter, A., & Wilson, J. (2021). Statewide implementation of automated writing evaluation: analyzing usage and associations with state test performance in grades 4-11. *Educational Technology Research & Development, 69*(3), 1557-1578. https://doi.org/10.1007/s11423-021-10004-9 | English Language |
| 60 | Prabhudesai, A., & Duong, T. N. B. (2019, 10-13 Dec. 2019). Automatic Short Answer Grading using Siamese Bidirectional LSTM Based Regression. *2019 IEEE International Conference on Engineering, Technology and Education (TALE), Yogyakarta, Indonesia*, 1-6. https://doi.org/10.1109/TALE48000.2019.9226026 | Computer Science |
| 61 | Prasain, B., & Bajaj, S. K. (2020, 25-27 Nov. 2020). Analysis of Algorithms in Automated Marking in Education: A Proposed Hybrid Algorithm. *2020 5th International Conference on Innovative Technologies in Intelligent Systems and Industrial Applications (CITISIA), Sydney, Australia*, 1-10. https://doi.org/10.1109/CITISIA50690.2020.9371783 | Multidisciplinary |
| 62 | Ratna, A. A. P., Ibrahim, I., & Purnamasari, P. D. (2017). Parallel Processing Design of Latent Semantic Analysis Based Essay Grading System with OpenMP. *Proceedings of the* | Indonesian Language |
| | *2017 International Conference on Computer Science and Artificial Intelligence, Jakarta, Indonesia.* https://doi.org/10.1145/3168390.3168401 | |
| 63 | Ratna, A. A. P., Purnamasari, P. D., Anandra, N. K., & Luhurkinanti, D. L. (2023). hybrid deep learning CNN-bidirectional LSTM and Manhattan distance for Japanese automated short answer grading: Use case in Japanese language studies. *Proceedings of the 8th International Conference on Communication and Information Processing, Beijing, China*. https://doi.org/10.1145/3571662.3571666 | Japanese Language |
| 64 | Rodriguez-Ruiz, J., Alvarez-Delgado, A., & Caratozzolo, P. (2021, 15-17 Dec. 2021). Use of Natural Language Processing (NLP) Tools to Assess Digital Literacy Skills. *2021 Machine Learning-Driven Digital Technologies for Educational Innovation Workshop, Monterrey, Mexico,* 1-8. https://doi.org/10.1109/IEEECONF53024.2021.9733779 | Engineering |
| 65 | Rosaria Re, M., Amenduni, F., De Medio, C., & Valente, M. (2019). How to use assessment data collected through writing activities to identify participants' critical thinking levels. *Journal of E-Learning & Knowledge Society, 15*(3), 117-132. https://doi.org/10.20368/1971-8829/1135051 | Higher Education Teachers |
| 66 | Rosenberg, J. M., & Krist, C. (2021). Combining machine learning and qualitative methods to elaborate students' ideas about the generality of their model-based explanations. Journal of Science Education & Technology, 30(2), 255-267. https://doi.org/10.1007/s10956-020-09862-4 | Science |
| 67 | Ruan, S., Jiang, L., Xu, J., Tham, B. J.-K., Qiu, Z., Zhu, Y., Murnane, E. L., Brunskill, E., & Landay, J. A. (2019). QuizBot: A Dialogue-based Adaptive Learning System for Factual Knowledge. *Proceedings of the 2019 CHI Conference on Human Factors in Computing Systems, Glasgow, Scotland UK*. https://doi.org/10.1145/3290605.3300587 | Multidisciplinary |
| 68 | Rupp, A. A., Casabianca, J. M., Krüger, M., Keller, S., & Köller, O. (2019). Automated Essay Scoring at Scale: A Case Study in Switzerland and Germany. *ETS Research Reports Series, 2019*(1), 1-23. https://doi.org/10.1002/ets2.12249 | English Language |
| 69 | Rus, V. (2018). Explanation-based Automated Assessment of Open Ended Learner Responses. *eLearning & Software for Education, 2*, 120-127. https://doi.org/10.12753/2066-026X-18-087 | Science |
| 70 | Saad, M. B., Jackowska-Strumiłło, L., & Bieniecki, W. (2018, 4-6 July 2018). ANN Based Evaluation of Student's Answers in E-Tests. *2018 11th International Conference on Human System Interaction (HSI), Gdansk, Poland*, 155-161. https://doi.org/ 10.1109/HSI.2018.8431340 | Computer Science |
| 71 | Saeed, M. M., & Gomaa, W. H. (2022, 8-9 May 2022). An Ensemble-Based Model to Improve the Accuracy of Automatic Short Answer Grading. *2022 2nd International Mobile, Intelligent, and Ubiquitous Computing Conference (MIUCC), Cairo, Egypt,* 337-342. https://doi.org/ 10.1109/MIUCC55081.2022.9781737 | Multidisciplinary |
| 72 | Sahu, A., & Bhowmick, P. K. (2020). Feature Engineering and Ensemble-Based Approach for Improving Automatic Short-Answer Grading Performance. *IEEE Transactions on Learning Technologies, 13*(1), 77-90. https://doi.org/10.1109/TLT.2019.2897997 | Computer Science |
| 73 | Santamaría Lancho, M., Hernández, M., Sánchez-Elvira Paniagua, Á., Luzón Encabo, J. M., & de Jorge-Botana, G. (2018). Using Semantic Technologies for Formative Assessment and Scoring in Large Courses and MOOCs. *Journal of Interactive Media in Education, 2018*(1), 1-10. https://doi.org/10.5334/jime.468 | Business |
| 74 | Sayeed, M. A., & Gupta, D. (2022, 14-16 Dec. 2022). Automate Descriptive Answer Grading using Reference based Models. *2022 OITS International Conference on Information Technology (OCIT), Bhubaneswar, India*, 262-267. https://doi.org/10.1109/OCIT56763.2022.00057 | Science |
| 75 | Seifried, J., Brandt, S., Kögler, K., Rausch, A., & Ehmke, T. (2020). The computer-based assessment of domain-specific problem-solving competence—A three-step scoring procedure. *Cogent Education, 7*(1), 1-23. https://doi.org/10.1080/2331186X.2020.1719571 | Vocational Education |
| 76 | Shermis, M. D., Mao, L., Mulholland, M., & Kieftenbeld, V. (2017). Use of Automated Scoring Features to Generate Hypotheses Regarding Language-Based DIF. *International Journal of Testing, 17*(4), 351-371. https://doi.org/10.1080/15305058.2017.1308949 | English Language |
| 77 | Shweta, P., & Adhiya, K. (2022, 17-19 June 2022). Comparative Study of Feature Engineering for Automated Short Answer Grading. *2022 IEEE World Conference on Applied Intelligence and Computing (AIC), Sonbhadra, India*, 594-597. https://doi.org/10.1109/AIC55036.2022.9848851 | Computer Science |
| 78 | Sinha, S. K., Yadav, S., & Verma, B. (2022, 29-31 March 2022). NLP-based Automatic Answer Evaluation. *2022 6th International Conference on Computing Methodologies and Communication (ICCMC)*, *Erode, India,* 807-811. https://doi.org/10.1109/ICCMC53470.2022.9754052 | None |
| 79 | Smith, A., Leeman-Munk, S., Shelton, A., Mott, B., Wiebe, E., & Lester, J. (2019). A Multimodal Assessment Framework for Integrating Student Writing and Drawing in | Science |
| | Elementary Science Learning. *IEEE Transactions on Learning Technologies, 12*(1), 3-15. https://doi.org/10.1109/TLT.2018.2799871 | |
| 80 | Somers, R., Cunningham-Nelson, S., & Boles, W. (2021). Applying natural language processing to automatically assess student conceptual understanding from textual responses. *Australasian Journal of Educational Technology, 37*(5), 98-115. https://doi.org/10.14742/ajet.7121 | Engineering |
| 81 | Sung, S. H., Li, C., Chen, G., Huang, X., Xie, C., Massicotte, J., & Shen, J. (2021). How Does Augmented Observation Facilitate Multimodal Representational Thinking? Applying Deep Learning to Decode Complex Student Construct. *Journal of Science Education & Technology, 30*(2), 210-226. https://doi.org/10.1007/s10956-020-09856-2 | Science |
| 82 | Tulu, C. N., Ozkaya, O., & Orhan, U. (2021). Automatic Short Answer Grading With SemSpace Sense Vectors and MaLSTM.. *IEEE Access, 9,* 19270-19280. https://doi.org/10.1109/ACCESS.2021.3054346 | Engineering |
| 83 | Uzun, K. (2018). Home-Grown Automated Essay Scoring in the Literature Classroom: A Solution for Managing the Crowd? *Contemporary Educational Technology, 9*(4), 423-436. https://doi.org/10.30935/cet.471024 | English Literature |
| 84 | Wang, C., Liu, X., Wang, L., Sun, Y., & Zhang, H. (2021). Automated Scoring of Chinese Grades 7–9 Students' Competence in Interpreting and Arguing from Evidence. *Journal of Science Education & Technology, 30(*2), 269-282. https://doi.org/10.1007/s10956-020-09859-z | Chinese Language |
| 85 | Wu, Y., Cao, X., & Tian, X. (2022). Short Answer Automatic Scoring Based on Multi-model Dynamic Collaboration. *2021 4th International Conference on Artificial Intelligence and Pattern Recognition, Xiamen, China*. https://doi.org/10.1145/3488933.3489013 | Multidisciplinary |
| 86 | Xia, J., & Zilles, C. (2023). Using Context-Free Grammars to Scaffold and Automate Feedback in Precise Mathematical Writing. *Proceedings of the 54th ACM Technical Symposium on Computer Science Education V. 1, Toronto ON, Canada*. https://doi.org/10.1145/3545945.3569728 | Computer Science |
| 87 | Xing, W., Lee, H.-S., & Shibani, A. (2020). Identifying patterns in students' scientific argumentation: content analysis through text mining using Latent Dirichlet Allocation. *Educational Technology Research & Development, 68*(5), 2185-2214. https://doi.org/10.1007/s11423-020-09761-w | Science |
| 88 | Yannakoudakis, H., Andersen, Ø. E., Geranpayeh, A., Briscoe, T., & Nicholls, D. (2018). Developing an automated writing placement system for ESL learners. *Applied Measurement in Education, 31*(3), 251-267. https://doi.org/10.1080/08957347.2018.1464447 | English Language |
| 89 | Ye, X., & Manoharan, S. (2018, 26-28 Sept. 2018). Machine Learning Techniques to Automate Scoring of Constructed-Response Type Assessments. *2018 28th EAEEIE Annual Conference (EAEEIE), Hafnarfjordur, Iceland,* 1-6. https://doi.org/10.1109/EAEEIE.2018.8534209 | Social Studies |
| 90 | Ye, X., & Manoharan, S. (2019). Providing automated grading and personalized feedback. *Proceedings of the International Conference on Artificial Intelligence, Information Processing and Cloud Computing, Sanya, China*. https://doi.org/10.1145/3371425.3371453 | Computer Science |
| 91 | Ye, X., & Manoharan, S. (2020). Marking Essays Automatically. *Proceedings of the 2020 4th International Conference on E-Education, E-Business and E-Technology*, *Shanghai, China*. https://doi.org/10.1145/3404649.3404657 | English Language |
| 92 | Yeruva, N., Venna, S., Indukuri, H., & Marreddy, M. (2023). Triplet Loss based Siamese Networks for Automatic Short Answer Grading. *Proceedings of the 14th Annual Meeting of the Forum for Information Retrieval Evaluation, Kolkata, India*. https://doi.org/10.1145/3574318.3574337 | Computer Science |
| 93 | Zhu, X., Wu, H., & Zhang, L. (2022). Automatic Short-Answer Grading via BERT-Based Deep Neural Networks. *IEEE Transactions on Learning Technologies, 15*(3), 364-375. https://doi.org/10.1109/TLT.2022.3175537 | Multidisciplinary |